\documentclass[twocolumn]{aastex631}

\usepackage{xspace}
\newcommand{\target}{Gl~229~B\xspace}
\newcommand{\host}{Gl~229A\xspace}
\newcommand{\targetsys}{Gl~229\xspace}

\newcommand{\mo}[1]{$\mathrm{#1}$}
\newcommand{\ej}{ExoJAX\xspace}
\usepackage{url}

\shorttitle{Subaru/IRD high-resolution spectroscopy of Gl~229~B}
\shortauthors{Kawashima et al.}
\graphicspath{{./}{figures/}}
\begin{document}

\title{Atmospheric retrieval of Subaru/IRD high-resolution spectrum of the archetype T-type brown dwarf Gl~229~B
\footnote{This research is based on data collected at the Subaru Telescope, which is operated by the National Astronomical Observatory of Japan. We are honored and grateful for the opportunity of observing the Universe from Maunakea, which has the cultural, historical, and natural significance in Hawaii.}}

\correspondingauthor{Yui Kawashima}
\email{ykawashima@kusastro.kyoto-u.ac.jp}

\author[0000-0003-3800-7518]{Yui Kawashima} %Y
\affiliation{Department of Astronomy, Graduate School of Science, Kyoto University, Kitashirakawa Oiwake-cho, Sakyo-ku, Kyoto 606-8502, Japan}
\affiliation{Frontier Research Institute for Interdisciplinary Sciences, Tohoku University, 6-3 Aramaki aza Aoba, Aoba-ku, Sendai, Miyagi 980-8578, Japan}
\affiliation{Department of Geophysics, Graduate School of Science, Tohoku University, 6-3 Aramaki aza Aoba, Aoba-ku, Sendai, Miyagi 980-8578, Japan}
\affiliation{Institute of Space and Astronautical Science, Japan Aerospace Exploration Agency, 3-1-1 Yoshinodai, Chuo-ku, Sagamihara, Kanagawa 252-5210, Japan}
\affiliation{Cluster for Pioneering Research, RIKEN, 2-1 Hirosawa, Wako, Saitama 351-0198, Japan}
%\affiliation{SRON Netherlands Institute for Space Research, Niels Bohrweg 4, 2333 CA Leiden, The Netherlands}

\author[0000-0003-3309-9134]{Hajime Kawahara} %Y
\affiliation{Institute of Space and Astronautical Science, Japan Aerospace Exploration Agency, 3-1-1 Yoshinodai, Chuo-ku, Sagamihara, Kanagawa 252-5210, Japan}
\affiliation{Department of Astronomy, Graduate School of Science, The University of Tokyo, 7-3-1 Hongo, Bunkyo-ku, Tokyo 113-0033, Japan}

\author[0000-0002-8607-358X]{Yui Kasagi} %Y
\affiliation{Institute of Space and Astronautical Science, Japan Aerospace Exploration Agency, 3-1-1 Yoshinodai, Chuo-ku, Sagamihara, Kanagawa 252-5210, Japan}
\author[0000-0001-6309-4380]{Hiroyuki Tako Ishikawa} %Y
\affiliation{Department of Physics and Astronomy, The University of Western Ontario, 1151 Richmond St, London, Ontario, N6A~3K7, Canada}
\author[0000-0003-1298-9699]{Kento Masuda} %Y
\affiliation{Department of Earth and Space Science, Graduate School of Science, Osaka University, 1-1 Machikaneyama-cho, Toyonaka, Osaka 560-0043, Japan}
\author[0000-0001-6181-3142]{Takayuki Kotani}
\affiliation{Astrobiology Center, 2-21-1 Osawa, Mitaka, Tokyo 181-8588, Japan}
\affiliation{Astronomical Science Program, The Graduate University for Advanced Studies, SOKENDAI, 2-21-1 Osawa, Mitaka, Tokyo 181-8588, Japan}
\affiliation{National Astronomical Observatory of Japan, 2-21-1 Osawa, Mitaka, Tokyo 181-8588, Japan}
\author[0000-0002-9294-1793]{Tomoyuki Kudo} %Y
\affiliation{Subaru Telescope, 650 N. Aohoku Place, Hilo, HI 96720, USA}
\author[0000-0003-3618-7535]{Teruyuki Hirano} %Y
\affiliation{Astrobiology Center, 2-21-1 Osawa, Mitaka, Tokyo 181-8588, Japan}
\affiliation{National Astronomical Observatory of Japan, 2-21-1 Osawa, Mitaka, Tokyo 181-8588, Japan}
\author[0000-0002-4677-9182]{Masayuki Kuzuhara}
\affiliation{Astrobiology Center, 2-21-1 Osawa, Mitaka, Tokyo 181-8588, Japan}
\author[0000-0003-4698-6285]{Stevanus K Nugroho}%Y
\affiliation{Astrobiology Center, 2-21-1 Osawa, Mitaka, Tokyo 181-8588, Japan}
\author[0000-0002-4881-3620]{John Livingston}%Y
\affiliation{Astrobiology Center, 2-21-1 Osawa, Mitaka, Tokyo 181-8588, Japan}
\affiliation{National Astronomical Observatory of Japan, 2-21-1 Osawa, Mitaka, Tokyo 181-8588, Japan}
\affiliation{Astronomical Science Program, The Graduate University for Advanced Studies, SOKENDAI, 2-21-1 Osawa, Mitaka, Tokyo 181-8588, Japan}
\author[0000-0002-7972-0216]{Hiroki Harakawa}
\author[0000-0001-9326-8134]{Jun Nishikawa}
\affiliation{National Astronomical Observatory of Japan, 2-21-1 Osawa, Mitaka, Tokyo 181-8588, Japan}
\affiliation{Astronomical Science Program, The Graduate University for Advanced Studies, SOKENDAI, 2-21-1 Osawa, Mitaka, Tokyo 181-8588, Japan}
\affiliation{Astrobiology Center, 2-21-1 Osawa, Mitaka, Tokyo 181-8588, Japan}
\author[0000-0002-5051-6027]{Masashi Omiya}
\affiliation{Astrobiology Center, 2-21-1 Osawa, Mitaka, Tokyo 181-8588, Japan}
\affiliation{National Astronomical Observatory of Japan, 2-21-1 Osawa, Mitaka, Tokyo 181-8588, Japan}
\author[0009-0006-9082-9171]{Takuya Takarada}
\affiliation{Astrobiology Center, 2-21-1 Osawa, Mitaka, Tokyo 181-8588, Japan}
\author[0000-0002-6510-0681]{Motohide Tamura} %Y
\affiliation{Department of Astronomy, Graduate School of Science, The University of Tokyo, 7-3-1 Hongo, Bunkyo-ku, Tokyo 113-0033, Japan}
\affiliation{Astrobiology Center, 2-21-1 Osawa, Mitaka, Tokyo 181-8588, Japan}
\affiliation{National Astronomical Observatory of Japan, 2-21-1 Osawa, Mitaka, Tokyo 181-8588, Japan}
\author{Akitoshi Ueda}
\affiliation{National Astronomical Observatory of Japan, 2-21-1 Osawa, Mitaka, Tokyo 181-8588, Japan}
\affiliation{Astronomical Science Program, The Graduate University for Advanced Studies, SOKENDAI, 2-21-1 Osawa, Mitaka, Tokyo 181-8588, Japan}

%% Note that the \and command from previous versions of AASTeX is now
%% depreciated in this version as it is no longer necessary. AASTeX 
%% automatically takes care of all commas and "and"s between authors names.

%% AASTeX 6.31 has the new \collaboration and \nocollaboration commands to
%% provide the collaboration status of a group of authors. These commands 
%% can be used either before or after the list of corresponding authors. The
%% argument for \collaboration is the collaboration identifier. Authors are
%% encouraged to surround collaboration identifiers with ()s. The 
%% \nocollaboration command takes no argument and exists to indicate that
%% the nearby authors are not part of surrounding collaborations.

%% Mark off the abstract in the ``abstract'' environment. 
\begin{abstract} %250 words
Brown dwarfs provide a unique opportunity to study atmospheres and their physical and chemical processes with high precision, especially in temperature ranges relevant to exoplanets.
In this study, we performed high-resolution ($R \sim 70,000$) spectroscopy using Subaru/IRD {($Y$, $J$, $H$-bands)} of {the T7.0p-type object} \target, the first discovered T-type brown dwarf, which orbits an M1V host star at a separation of 33~au.
We conducted atmospheric retrieval on the reduced $H$-band spectrum using the high-resolution spectrum model compatible with automatic differentiation and GPU, \ej.
In contrast to previous retrieval studies on medium-resolution spectra, we obtained a C/O ratio consistent with that of the host star, aligning with the expected formation process for such a massive brown dwarf.
Additionally, based on the strong constraint on temperature from the high-resolution spectrum and previously measured photometric magnitude, our analysis indicates that \target is a binary, which was also proposed by \citet{2021AJ....162..301B} {and recently confirmed by \citet{2024Natur.634.1070X}}.
Finally, we validated current molecular line lists by leveraging the obtained high-resolution, {high signal-to-noise ratio} spectrum of this warm ($\sim 900$~K) atmosphere.
This study highlights the importance of observing companion brown dwarfs as benchmark objects for establishing characterization techniques for low-mass objects and enhancing our understanding of their atmospheres, given the wealth of available information and the relative ease of observation.
\end{abstract}

%% Keywords should appear after the \end{abstract} command. 
%% The AAS Journals now uses Unified Astronomy Thesaurus concepts:
%% https://astrothesaurus.org
%% You will be asked to selected these concepts during the submission process
%% but this old "keyword" functionality is maintained in case authors want
%% to include these concepts in their preprints.
%\keywords{Classical Novae (251) --- Ultraviolet astronomy(1736) --- History of astronomy(1868) --- Interdisciplinary astronomy(804)}
% alphabetical order

%% From the front matter, we move on to the body of the paper.
%% Sections are demarcated by \section and \subsection, respectively.
%% Observe the use of the LaTeX \label
%% command after the \subsection to give a symbolic KEY to the
%% subsection for cross-referencing in a \ref command.
%% You can use LaTeX's \ref and \label commands to keep track of
%% cross-references to sections, equations, tables, and figures.
%% That way, if you change the order of any elements, LaTeX will
%% automatically renumber them.
%%
%% We recommend that authors also use the natbib \citep
%% and \citet commands to identify citations.  The citations are
%% tied to the reference list via symbolic KEYs. The KEY corresponds
%% to the KEY in the \bibitem in the reference list below. 

\section{Introduction} \label{sec:intro}
Brown dwarfs are objects that fill the gap between planets and stars.
Understanding the atmospheres of brown dwarfs is essential for revealing their formation and evolution processes, thereby obtaining a seamless view of objects ranging from planets to low-mass stars.
Brown dwarfs also share many physical and chemical atmospheric processes with planets inside and outside our solar system, such as chemical reactions {\citep[e.g.,][]{2014ApJ...797...41Z}}, atmospheric circulation {\citep[e.g.,][]{2016Natur.533..330S}}, and cloud formation {\citep[e.g.,][]{2021JGRE..12606655G}}, making them crucial for comprehensively understanding such substellar atmospheres.
%Thus, the atmospheric spectra of brown dwarfs, which can be observed with relatively high precision, serve as excellent templates for the current and near-future atmospheric characterization of exoplanets that have similar temperatures and compositions.

By resolving individual atomic and molecular absorption line profiles in brown dwarfs and planets, high-resolution ($R \gtrsim 10,000$) spectroscopy can provide unique, in-depth information.
%, which is different from what can be obtained from low-/medium-resolution spectroscopy.
High-resolution spectra are sensitive to temperature through line strength ratios, the presence of minor chemical species{, including metal-bearing molecules such as TiO and Ti} \citep[e.g.,][]{2017AJ....154..221N, 2021AJ....161..153I} {and radical species often found in upper atmospheres, such as OH \citep{2021ApJ...910L...9N}}, and dynamical information, such as atmospheric winds and rotation \citep{2010Natur.465.1049S, 2012ApJ...751..117M}.

{However,} high-resolution spectroscopic observations of brown dwarfs{, particularly of their} atmospheres, have been relatively scarce, {and most of these observations were carried out at $R \lesssim 50,000$}. They include the Gemini/Phoenix observation of $\epsilon$ INDI Ba \citep[wavelength ranges of 1.553--1.559 and 2.308--2.317 $\mu$m, $R \gtrsim 50,000$;][]{2003ApJ...599L.107S}, Keck II/NIRSPEC observations of dozens of brown dwarfs in the $J$-band \citep[1.165--1.324~$\mu$m, $R \sim 20,000$;][]{2007ApJ...658.1217M, 2010ApJS..186...63R}, VLT/UVES observations of three L dwarfs and one T dwarf binary system \citep[0.64--1.02~$\mu$m, $R \sim 33,000$;][]{2007A&A...473..245R}, VLT/CRIRES observations of Luhman~16AB \citep[2.288--2.345~$\mu$m, $R \sim 100,000$;][]{2014Natur.505..654C}, and the recent Keck/KPIC/NIRSPEC observation of a binary L-type brown dwarf HR~7672B \citep[1.952--2.501~$\mu$m, $R \sim 35,000$;][]{2022AJ....163..189W} and Gemini South/IGRINS observation of a T6 dwarf \citep[1.46--2.48~$\mu$m, $R \sim 45,000$;][]{2022MNRAS.514.3160T}.

While high-resolution spectroscopy has been extensively used for exoplanet atmospheric characterization, one of the most significant issues is the validity of molecular line lists at such high temperatures ($\gtrsim$ 1,000~K).
The detection of molecular species through high-resolution spectroscopy of exoplanet atmospheres usually relies on a cross-correlation technique with the use of template spectra to boost their relatively low signal.
To make use of this technique, precise knowledge of the strength and central wavelength of each absorption line is essential.
While molecular line lists suitable for such high-temperature environments have been actively developed over the {past decade} by projects like HITEMP \citep{2010JQSRT.111.2139R} and ExoMol \citep{2016JMoSp.327...73T}, observational validation of those line lists has been limited.
In this context, 
%Under such a circumstance, as demonstrated by
%likewise, 
%\citet{2022MNRAS.514.3160T} and \citet{ 2023ApJ...953..170H}, 
{because brown dwarfs typically exhibit higher signal-to-noise ratios than exoplanets, their}
% high-precision 
high/medium-resolution spectra provide us with a unique opportunity to examine molecular line lists that are also used for exoplanet atmospheric characterization, as reported in \citet{2022MNRAS.514.3160T} and \citet{2023ApJ...953..170H}.
%Such molecular line list validation is essential for the observation of exoplanet atmospheres, given its typically lower achievable signal-to-noise ratios.

While the {signal-to-noise ratios currently achievable for}
% observational precision of 
high-resolution spectra %for 
{of} exoplanets is relatively limited, there are forthcoming 30~m-class telescopes, such as the Extremely Large Telescope (ELT) and Thirty-Meter Telescope (TMT).
The achievable signal-to-noise ratio with such 30~m-class telescopes should enable direct comparison of observed and model spectra as currently done for brown dwarfs, namely without the help of the cross-correlation technique.
Therefore, establishing analysis techniques using already available {high signal-to-noise ratio}, high-resolution spectra of brown dwarf atmospheres is essential for paving the way for future exoplanet characterization with next-generation telescopes.
% Furthermore, the achievable signal-to-noise ratio of atmospheric spectra for brown dwarfs with current 10~m-class ground-based telescopes should be approximately comparable to that for exoplanets with forthcoming 30~m-class telescopes, such as the Extremely Large Telescope (ELT) and Thirty-Meter Telescope (TMT).
% While the cross-correlation technique with template spectra is commonly employed for high-resolution spectroscopic characterization of exoplanet atmospheres, direct comparison of observed and model spectra is standard for brown dwarfs.
% %As indicated by this fact, 
% This is because the spectral analysis method significantly depends on the achievable observational signal-to-noise ratio.
% Therefore, establishing characterization techniques using already available high-precision, high-resolution spectra of brown dwarf atmospheres is essential for paving the way for future exoplanet characterization with next-generation telescopes, which is expected to realize a signal-to-noise ratio similar to that for the current observations of brown dwarfs.

Compared to the hotter L-type brown dwarfs, T-type brown dwarfs that are bright enough for detailed analysis using high-resolution spectroscopy are rare.
One of the valuable examples is 
%Among the known brown dwarfs, 
\target, which was the first T-type brown dwarf discovered \citep{1995Natur.378..463N}. While it has a spectral type of T7.0p \citep{2006ApJ...637.1067B} and relatively low effective temperature \citep[$\sim 800$--$900$~K;][]{2022ApJ...940..164C}, its proximity {\citep[5.7612~pc;][]{2020yCat.1350....0G}} makes it exceptionally bright \citep[$H=14.36${~mag};][]{2012ApJ...752...56F}.
This has led to the intensive investigation of its atmosphere compared to other T-type dwarfs discovered later, and it has served as the benchmark of T-type dwarfs.

Soon after the discovery of \target, \citet{1995Sci...270.1478O} observed its near-infrared spectrum and revealed the absorption features of \mo{CH_4}, followed by the detection of CO at 4.5~$\mu$m by \citet{1997ApJ...489L..87N}. \citet{2000ApJ...541..374S} reported the detection of $H$- and $K$-band features of \mo{NH_3}, but only tentatively due to their overlaps with the strong features of \mo{H_2O} and \mo{CH_4}.
They also hinted at the presence of \mo{H_2S}, though it was unclear since the increase in \mo{NH_3} abundance could compensate for its presence.

Recently, \citet{2022ApJ...935..107H} and \citet{2022ApJ...940..164C} performed a retrieval for the compilation of the previously measured medium-resolution spectrum \citep{1996ApJ...467L.101G, 1998ApJ...492L.181S, 1998ApJ...502..932O, 1997ApJ...489L..87N} and constrained the abundances of several chemical species in the atmosphere.
They both reported a high C/O ratio of $\gtrsim 1$, which is much larger than the value for the host star \citep[$0.68 \pm 0.12$;][]{2015AJ....150...53N}.
%They also performed the retrieval setting the object mass as a free parameter and estimated it to be $41.6 \pm 3.3 M_\mathrm{J}$ (their fiducial case) and $50^{+12}_{-9} M_\mathrm{J}$, respectively.

Further, \citet{2021AJ....162..301B} recently presented a precise dynamical mass measurement of $71.4 \pm 0.6 M_\mathrm{J}$ from combined observations of astrometry and radial velocity.
They found a discrepancy between the measured dynamical mass and the mass predicted by the evolutionary models, and proposed the scenario that \target is an unresolved binary.
%to explain its exceptionally low luminosity for the large mass they measured.
{Recently, \citet{2024ApJ...974L..30W} and \citet{2024Natur.634.1070X} \footnote{These papers were published while the present manuscript was under review.} independently resolved Gl 229 B into two components, using radial velocity measurements and a combination of GRAVITY interferometry and radial velocity, respectively.}

To revisit the anomalous elemental abundances that are inconsistent with the host star and explore the binary scenario, a high-resolution spectrum of \target containing a wealth of information is useful.
In this paper, we present the high-resolution spectroscopic observation of the archetype T-type brown dwarf \target, using Subaru/IRD \citep{2012SPIE.8446E..1TT, 2018SPIE10702E..11K}, and the retrieval results obtained by applying the high-resolution spectrum model ExoJAX \citep{2022ApJS..258...31K, exojax2} to the IRD spectrum.

The rest of this paper is organized as follows. In Section~\ref{sec:obs}, we describe our observations and data reduction, followed by the description of the spectral retrieval method in Section~\ref{sec:retrieval}.
In Section~\ref{sec:result}, we present our results of the spectral retrieval and molecular line list validation by utilizing the obtained {high signal-to-noise ratio}, high-resolution spectrum of a T-type brown dwarf.
In Section~\ref{sec:discussion}, we compare our results with previous studies and discuss several points that should be addressed in future studies.
Finally, we summarize this paper in Section~\ref{sec:summary}.

\section{Observations and Data reduction} \label{sec:obs}
\subsection{Subaru/IRD Observation}
We observed \targetsys on the first half night of 2021, January 31 (Universal Time) using the InfraRed Doppler instrument \citep[IRD;][]{2012SPIE.8446E..1TT, 2018SPIE10702E..11K} and the Adaptive Optics system AO188 mounted on the 8.2~m Subaru telescope (ID: S20B-105, PI: Y. Kawashima).
IRD {spans} three wavelength bands, $Y$$J$$H$ (0.97--1.75~$\mu$m), with $R \sim 70,000$ {distributed across 72 spectral orders}.
{The airmass at the start of the observations was 1.342, increasing to 1.588 by the final exposure.}

\begin{figure}[ht!]
\plotone{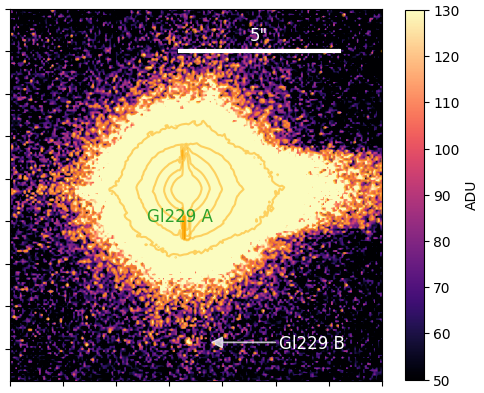}
\caption{Obtained image of the \targetsys system (north at the top and east at the right) taken by the IRD Fiber Injection Module (FIM) acquisition CCD camera (FWHM wavelength range of $\sim 840$--$990$~nm). 
\label{fig:fim}}
\end{figure}

We performed the observations as follows. We first took images of the system using a CCD acquisition camera on the Fiber Injection Module (FIM) of IRD (see Figure~\ref{fig:fim}), which covers a F{u}ll Width at Half Maximum (FWHM) wavelength range of $\sim 840$--$990$~nm. We identified the target Gl~229~B at the angular separation and the position angle of $4.89 \arcsec$ and $179.5^\circ$ respectively, which are consistent with those extrapolated from Table~1 of \citet{2020AJ....160..196B}.
Then we injected the light from Gl~229~B to a multi-mode fiber of IRD, which has a diameter of 0.48\arcsec, in order to obtain spectrum with the spectrograph.

We used the host star \host as a natural guide star and used a ThAr ramp for wavelength calibration.
We continuously observed the target for six frames from 22:06-00:08 (HST) with an exposure time of 1,200~s per frame.
We achieved a signal-to-noise ratio of $\sim 30$--$50$ (combined six frames) for the $H$-band spectral orders 56--66, which we use for the spectral retrieval in this study.
Given the wide separation of the target from the host star, we consider the speckle contamination to be negligible.
%Airmass increases from 1.342 to 1.577 during the observation.
After observing \target, we also observed its host star \host for five frames with an exposure time of 60~s.
{These observations were conducted to determine the metallicity of the host star, which serves as a reference for characterizing the atmosphere of \target.}

%軌道予測されたpositionにいた点源をmulti mode fiberにinject
%separationを書く
%4秒離れて直径0.48秒
% kotani-san astrometric position from FIM image
% Aをnatural guide star
% Hバンドピークでの6枚合わせたもののphoton count

\subsection{Data Reduction} \label{sec:reduction}
We reduced the data with the Python pipeline for Subaru/IRD, PyIRD\footnote{\url{https://github.com/prvjapan/pyird}} \citep[][release version 0.3.0]{pyIRD}.
Using the six frames of the target data normalized by the flat data, we first derived the median 1D normalized spectrum.
We masked the pixels whose counts in the flat data were less than half of the median values of the respective spectral orders and those within 5 km/s wavelength distance from the OH telluric lines, which we took from \citet{2000A&A...354.1134R}, and sky emission lines observed by IRD.
We corrected the telluric transmittance by dividing by the theoretical transmission spectra, as \citet{2020PASJ...72...93H} generated with the Line-By-Line Radiative Transfer Model \citep[LBLRTM;][]{2005JQSRT..91..233C} for the grids of the precipitable water vapor content and target airmass.
We then applied the heliocentric radial velocity 
correction calculated with {\tt Astropy} \citep{2013A&A...558A..33A, 2018AJ....156..123A}.
We only corrected the heliocentric radial velocity, so the radial velocity value we retrieve in \S~\ref{sec:result} is the absolute value for the target, not the value relative to the host star \host.
To remove outliers, we masked the pixels whose counts were more than ten times the median values of the respective spectral orders or were negative.
Then, we further masked the neighboring 5 pixels if the count deviated more than 3$\sigma$ from the value of the detrended spectrum calculated by the median-filtering with the neighboring 11 pixels.
At this stage, the obtained spectrum still included the wavelength dependence of the flat lamp.
We assumed this wavelength dependence as that of the black body radiation of a somewhat uncertain temperature of the flat lamp.
To estimate this temperature, we compared the moving averaged IRD spectrum of a telluric standard star HR~2297 (B8IIIn) for the wavelength range longer than 1.5~$\mu$m, which is the region considered for the spectral retrieval in this study, and the spectrum of a similar spectral type (B9III) taken from the Pickles Atlas spectral library\footnote{\url{https://www.stsci.edu/hst/instrumentation/reference-data-for-calibration-and-tools/astronomical-catalogs/pickles-atlas}} (filename of \verb|"pickles_uk_64"|).
The fitted temperature was 1243~K and we multiplied by the black body spectrum of this temperature.
{The reduced spectrum is shown in Figures~\ref{fig:spec} and \ref{fig:spec_all}.}

{For our retrieval analysis, we use the normalized spectrum; we did not perform flux calibration. Instead, we incorporate the literature H-band magnitude as additional information to constrain the absolute flux. To estimate the flux uncertainties, we account for both photon noise and jitter noise. Jitter noise encompasses all remaining noise sources beyond photon noise and is treated as a free parameter in the retrieval. See \S~\ref{sec:obs_data} for details.
}

% Finally, we converted the observed relative flux to the absolute one using the previously-measured medium-resolution spectrum of \target as a reference, which we obtain from the L and T type dwarf archive\footnote{https://staff.gemini.edu/~sleggett/LTdata.html}.
% That spectrum is a compilation of HST/STIS spectrum \citep{1998ApJ...492L.181S} calibrated with the photometry observation of \citet{1998AJ....115.2579G}, $J$$H$$K$-band spectrum \citep{1996ApJ...467L.101G} with the flux recalibrated by \citet{1999ApJ...517L.139L}, $L$-band spectrum \citep{1998ApJ...502..932O}, and 5~$\mu$m spectrum \citep{1997ApJ...489L..87N} calibrated with the photometry observation of \citet{2004AJ....127.3516G}.
% %no clear res. info in the papers
% The spectrum in the wavelength range of our observation (0.97--1.75~$\mu$m) comes from \citet{1996ApJ...467L.101G, 1998ApJ...492L.181S, 1999ApJ...517L.139L}.
% We fit our observed spectrum of each spectral order to the medium-resolution spectrum so that the residual sum of the square yields the minimum.
% Figure~\ref{fig:fit} shows the resultant Subaru/IRD spectrum of \target along with the previously-measured medium-resolution spectrum used as a reference.

% \begin{figure*}[ht!]
% \plotone{figures/fit.pdf}
% \caption{The high-resolution spectrum of \target observed by Subaru/IRD (blue line) along with the preciously-observed medium-resolution spectrum \citep[orange line;][]{1996ApJ...467L.101G, 1998ApJ...492L.181S, 1999ApJ...517L.139L}, which we used as a reference to derive the absolute flux.
% \label{fig:fit}}
% \end{figure*}

\subsection{Metallicity of \host
%Analysis of the Host Star's Spectrum
} \label{sec:host}
%\ch{Check with Ishikawa-kun} -> done
%To better understand the observed target and system, 
{For the IRD spectrum of the host star \host, we performed data reduction following the procedure described in Section 2 of \citet{2022AJ....163...72I}, excluding the steps of IP deconvolution and telluric correction. These steps, which are essential for precision radial velocity measurements, were omitted in this study because they could degrade the signal-to-noise ratio, and were not essential for the scientific goal of determining the stellar metallicity.
We then analyzed the reduced spectrum}
in the same way as \citet{2020PASJ...72..102I, 2022AJ....163...72I} to derive its metallicity using 36 atomic lines {of eight elements (Na, Mg, Si, Ca, Ti, Cr, Mn, and Fe)}. 
{First, for each absorption line, we measured the abundance of the corresponding element by comparing the observed equivalent width (EW) with that of a synthetic spectrum computed using the model of \citet{1978A&A....62...29T}. We iteratively adjusted the elemental abundance until the calculated EW matched the observed EW.
Finally, we determined the abundance of each element by averaging the values obtained from its individual lines.
For a detailed description of this procedure, see Section 3 of \citet{2022AJ....163...72I}.}
{Note that while they adopted the effective temperature, which is essential for the abundance analysis, from the TESS Input Catalog, we determined it in this study based on 30 selected iron hydride (FeH) lines with clear line profiles in the $Y$-band, ensuring consistency between the derived temperature and metallicity. This method was first adopted in \citet{2022AJ....163..298M}, where the detailed procedure is described in their Section 3.1.1.}

From the weighted average of abundances of the eight elements, %Na, Mg, Si, Ca, Ti, Cr, Mn, and Fe, 
we calculated the metallicity of \host to be $[\mathrm{Fe}/\mathrm{H}] = 0.30 \pm 0.13$~dex. %(, where the weights were determined based on the uncertainties in the abundance measurements, )
This value is close to the reported abundances of C and O by \citet{2015AJ....150...53N}, $0.27 \pm 0.08$~dex and $0.21 \pm 0.04$~dex, respectively \footnote{We used the solar elemental abundance of \citet{2021A&A...653A.141A} to derive the abundances of carbon and oxygen relative to the solar values.}.

\section{Spectral Retrieval} \label{sec:retrieval}
% ExoJAXはdifferentiableなのでHMCができる
We performed spectral retrieval for the reduced Subaru/IRD data using our recently-developed high-resolution spectrum code for exoplanets and brown dwarfs, \ej\footnote{\url{https://github.com/HajimeKawahara/exojax}} \citep[][release version 1.5]{2022ApJS..258...31K, exojax2}.
\ej is an end-to-end simulation code from molecular line list databases such as HITRAN \citep{2022JQSRT.27707949G}/HITEMP \citep{2010JQSRT.111.2139R} and ExoMol \citep{2024JQSRT.32609083T} to a final observable spectrum model with consideration of the instrument response.
Also, \ej is compatible with automatic differentiation \citep{10.1145/355586.364791, 2015arXiv150205767G} and GPU with the use of JAX \citep{jax2018github}.
Thanks to this capability, it can perform Hamiltonian Monte Carlo (HMC) calculations using recent probabilistic programming languages such as NumPyro \citep{2019arXiv191211554P, JMLR:v20:18-403}.
In this study, we performed sampling with HMC and No U-Turn Sampler (NUTS) implemented in NumPyro.
For a detailed description of \ej, see \citet{2022ApJS..258...31K} and \citet{exojax2}.

%Note that d
Due to the enormously expensive computation of a high-resolution spectral retrieval, we limited the IRD spectral orders considered to 56--66, giving the wavelength range 14,676---16,470~$\mathrm{\AA}$.
This region corresponds to the wavelength range where late-T spectral-type objects like the target emit the highest radiation in the $H$-band.
Below, we describe our retrieval method with \ej.

\begin{deluxetable*}{lcccc}
\tablenum{1}
\tablecaption{Retrieval parameters of \target \label{tab:params}}
\tablewidth{0pt}
\tablehead{
\colhead{Parameter} & \colhead{Symbol} & \colhead{Prior} & \colhead{Unit} & \colhead{Reference}
}
\startdata
Mass & $M$ & $\mathcal{N}_t (71.4, 0.6, \mathrm{low = 1.0})$/$\mathcal{U} (1, 150)$ & $M_\mathrm{J}$ & \citet{2021AJ....162..301B} \\
Gravity & $\log_{10} g$ & $\mathcal{U} (-4.0, -6.0)$ & cgs for $g$ & \\
Volume mixing ratio of species $i$ & $\log_{10} x_i$ & $\mathcal{U} (-10, 0)$ & & \\
Temperature at 1~bar & $T_0$ & $\mathcal{U} (500, 1500)$ & $\mathrm{K}$ & \\
Temperature gradient & $\alpha$ & $\mathcal{U} (0.0, 0.2)$ & & \\
Radial velocity & RV & $\mathcal{U} ({-20}, {20})$ & $\mathrm{km}$ $\mathrm{s}^{-1}$& \\
Projected rotational velocity & $v \sin{i}$ & $\mathcal{U} (0, 40)$ & $\mathrm{km}$ $\mathrm{s}^{-1}$& \\
Jitter noise & $\sigma_\mathrm{j}$ & $\mathcal{E} (10)$& & \\
\enddata
\tablecomments{$\mathcal{U} (a, b)$ denotes the uniform distribution between $a$ and $b$, while $\mathcal{E} (\lambda)$ is the exponential distribution with a rate parameter of $\lambda$.
In the mass-constrained retrieval, we adopted a truncated normal distribution $\mathcal{N}_t$ with a mean of $71.4$, standard deviation of $0.6$, and lower bound of $1.0$ to avoid negative mass.
In the mass-free retrieval, we used a uniform distribution of $\mathcal{U} (1, 150)$.}
\end{deluxetable*}

\subsection{Opacity}
%absorption
For the molecular line absorption, we considered those for \mo{H_2O}, \mo{CH_4}, and \mo{NH_3}, which are the major chemical species that can contribute to the $H$-band spectrum of T-type brown dwarfs {\citep[e.g.,][]{2000ApJ...541..374S, 2005ARA&A..43..195K}}.
We only considered their main isotopologues, namely $^{1}\mathrm{H}_2^{16}\mathrm{O}$, $^{12}\mathrm{C}^{1}\mathrm{H}_4$, and $^{14}\mathrm{N}^{1}\mathrm{H}_3$.
%Note that 
{We excluded CO, Na, and K, which were included in previous medium-resolution retrievals of this object \citep{2022ApJ...935..107H, 2022ApJ...940..164C}, because their contributions are almost negligible in our wavelength range. Also, }
while the presence of \mo{H_2S} was recently reported by \citet{2022MNRAS.514.3160T} for another T-type brown dwarf, we opted to ignore it in this study.
This is because the considered wavelength range includes only one relatively strong feature of \mo{H_2S} at 1.59~$\mu$m, while there are numerous absorption features of \mo{H_2O}, \mo{CH_4}, and \mo{NH_3}.
Indeed, when we attempted spectral retrieval including \mo{H_2S}, the HMC simulation suffered from making its abundance converge.
%This is probably because there is only one tentative feature of \mo{H_2S} at 1.59~$\mu$m, while there are numerous absorption features of \mo{H_2O}, \mo{CH_4}, and \mo{NH_3} in the considered wavelength range.

We used the line lists of POKAZATEL \citep{2018MNRAS.480.2597P} and CoYuTe \citep{2015JQSRT.161..117A, 2019MNRAS.490.4638C} from ExoMol \citep{2016JMoSp.327...73T} for \mo{H_2O} and \mo{NH_3}, respectively.
For \mo{CH_4}, we used the HITEMP line list \citep{2010JQSRT.111.2139R, 2020ApJS..247...55H}, as it was recently proven to be more accurate for the analyses of the observed spectra of other T-type brown dwarfs \citep{2022MNRAS.514.3160T, 2023ApJ...953..170H}.
We explore the validity of the current molecular line lists in \S~\ref{linelist}.

We also considered the collision-induced absorption (CIA) of \mo{H_2}--\mo{H_2} and \mo{H_2}--\mo{He}, taking those data from HITRAN \citep{2019Icar..328..160K}.
In this study, we ignored the effect of clouds, as recent medium-resolution spectral retrieval for this object showed that their inclusion is unnecessary \citep{2022ApJ...940..164C}.

\subsection{Atmospheric Structure}
% abundance
We retrieved the volume mixing ratios of the three chemical species \mo{H_2O}, \mo{CH_4}, and \mo{NH_3}, assuming vertically constant abundances for simplicity.
We assumed that the remaining mass is composed of \mo{H_2} and \mo{He} with a relative volume mixing ratio of 6:1, similar to the solar elemental abundance ratio \citep{2021A&A...653A.141A}.

%T-P
For the temperature--pressure structure, we used a power-law profile, following \citet{2022ApJS..258...31K}.
\begin{equation}
    T = T_0 \left( \frac{P}{1 \ \mathrm{bar}} \right)^{\alpha},
\end{equation}
where $T$ and $P$ are the temperature and pressure.
$T_0$ and $\alpha$, the retrieval parameters, are a reference temperature at 1~bar pressure and a parameter for controlling the temperature gradient, respectively.
Due to the expensive calculation of high-resolution spectral retrieval, we leave employing more flexible profiles for future studies.

The radius and gravity are both needed to simulate the spectrum.
{Although} any two of the three parameters, radius, gravity, and mass, {could} be {treated as free} parameters, we chose to retrieve mass and gravity {and then compute the radius, since it is straightforward to assign priors to mass and gravity}.
For the mass, in our fiducial retrieval, we used the dynamical mass measurement of $71.4 \pm 0.6 M_\mathrm{J}$ from \citet{2021AJ....162..301B} as the prior information.
In this paper, we refer to this retrieval as ``mass-constrained retrieval.''
To explore the possibility of constraining the object mass from the observed high-resolution spectrum, we also performed ``mass-free retrieval,'' imposing a uniform prior probability distribution of $1$--$150$~$M_\mathrm{J}$.
{In both retrievals, we used a uniform prior $\mathcal{U} (-4.0, -6.0)$ on gravity $\log_{10} g$~(cgs).}

\subsection{Radiative Transfer and Simulated Spectrum}
We simulated the high-resolution spectrum with the pure-absorption (i.e., without scattering) intensity-based (8 streams) radiative transfer scheme of \ej.
The details of this scheme can be found at \citet{exojax2}.
To simulate the final observable spectrum, we considered the wavelength shift and line broadening by treating the radial velocity and projected rotational velocity as retrieval parameters, {with uniform priors $\mathcal{U} (-20, 20)$ and $\mathcal{U} (0, 40)$~$\mathrm{km}$/$\mathrm{s}$, respectively (see Table~\ref{tab:params}). We also accounted for} the instrumental response of IRD.
For the details of these procedures, see Section~4 of \citet{2022ApJS..258...31K}.
We assumed the two limb-darkening parameters as $u_1 = 0$ and $u_2 = 0$.

% details
We set %the resolution of 
the relevant parameters required for the simulation so that the uncertainty of the final spectrum model to be compared with the observed IRD data was $\lesssim 1\%$:
We only considered the lines with intensities $> 10^{-30}$~$\mathrm{cm}^{2}$~$\mathrm{cm}^{-1}$ at 1000~$\mathrm{K}$.
We used 200 atmospheric layers evenly spaced in a logarithmic manner in the pressure range $10^{-3}$--$10^2$~bar. For the calculation of the original spectrum (the spectrum before the rotational broadening and instrumental response are applied), we adopted a spectral resolution of $R = 700,000$, ten times higher than the IRD resolution \citep[$R \sim 70,000$;][]{2018SPIE10702E..11K}.
We used the 32-bit mode for the JAX calculation rather than the 64-bit mode as we have confirmed its validity.

\subsection{Comparison with the Observed Data} \label{sec:obs_data}
%flux
Since the absolute flux is not measurable from our Subaru/IRD high-resolution spectroscopic observation, at each step inside HMC, we found the scaling parameter $a$ for which the simulated spectrum
%best 
matches the Subaru/IRD data.
%We minimized $\chi^2$ written as follows, using the Adam optimization implemented in JAXopt \citep{deepmind2020jax, 2021arXiv210515183B}:
We obtained the value of $a$ that minimizes $\chi^2$ written as follows, by calculating its derivative with respect to $a$:
\begin{equation} \label{eq:scaling}
    \chi^2 = \sum_i^\mathcal{N} \frac{\left (d_i - a m_i \right) ^2}{{\sigma_{\mathrm{p}, i}}^2 + {\sigma_j}^2},
\end{equation}
where $i$ and $\mathcal{N}$ denote the wavenumber point and its total number for the observed data.
$d_i$ and $m_i$ are the normalized observed flux and model flux at the wavenumber point $i$, respectively.
$\sigma_\mathrm{j}$ is the jitter noise, for which we attributed any remaining noises other than the photon noise $\sigma_{\mathrm{p}, i}$ and treated it as a retrieval parameter.

In addition to the Subaru/IRD data, we used the observed $H$-band magnitude of $d_H \pm \sigma_H = 14.36 \pm 0.05${~mag} \citep{2012ApJ...752...56F} to constrain the absolute flux. For the calculation of the model $H$-band magnitude $m_H$, we considered the transmission curve for the MKO NSFCam {$H$-band} filter taken from the SVO Filter Profile Service \citep{2012ivoa.rept.1015R, 2020sea..confE.182R}\footnote{\url{http://svo2.cab.inta-csic.es/theory/fps/}} and use a target system distance of {5.7612} pc \citep{2020yCat.1350....0G}.

{We emphasize that even for this photometric magnitude calculation, we modeled the spectrum at a resolution of $R = 100,000$, as we confirmed that such high resolution is necessary to keep the uncertainty in $m_H$ below $\sim$ 10\% of the observed uncertainty of 0.05 (i.e., 0.005), as demonstrated in Figure~\ref{fig:Hmag} in Appendix~\ref{sec:res}.
See Appendix~\ref{sec:res} for further details.}

% HMC
Assuming that the observational noise for both the IRD high-resolution spectrum and $H$-band magnitude obeys an independent normal distribution, the likelihood can be modeled as
\begin{eqnarray} \label{eq:likelihood}
    \mathcal{L} = &&\frac{1}{\sqrt{2 \pi \sigma_H^2}} \exp{\left( - \frac{\left (d_H - m_H \right) ^2}{2 \sigma_H^2} \right)} \nonumber \\
    &&\prod_i \frac{1}{\sqrt{2 \pi \left(\sigma_{\mathrm{p}, i}^2 + \sigma_j^2 \right)}} \exp{\left( - \frac{\left (d_i - a m_i \right) ^2}{2 \left( \sigma_{\mathrm{p}, i}^2 + \sigma_j^2 \right)} \right)}.
\end{eqnarray}
Note that in the above equation, the scaling parameter $a$ was obtained deterministically as described above.
The retrieval parameters considered are summarized in Table~\ref{tab:params}, together with the adopted prior probability distributions.
For the sampling with HMC, we used 500 warmup steps and 1000 samples.
The values of the convergence diagnostic $\hat{R}$ for all the parameters were below {1.02} and {1.01} for the mass-constrained and mass-free retrievals, respectively, indicating that the simulations converged well. 
The total computational times with an NVIDIA A100 80 GB PCIe GPU were {195} and {398}~hr for the mass-constrained and mass-free retrievals, respectively.

\section{Results} \label{sec:result}
As described in the previous section, in this paper, we performed two retrievals with different assumptions for the prior probability distribution of the object mass.
In \S~\ref{res:mass-const}, we first present the results of the mass-constrained retrieval, in which we used the dynamical mass measurement of $71.4 \pm 0.6 M_\mathrm{J}$ from \citet{2021AJ....162..301B}.
In \S~\ref{res:mass-free}, we show the results of the mass-free retrieval, in which we imposed a uniform prior of $1$--$150$~$M_\mathrm{J}$ to explore the possibility of constraining the mass from the high-resolution spectrum.
In addition, we examined the current molecular line lists by utilizing the observed {high signal-to-noise ratio}, high-resolution spectrum of a late T-type brown dwarf in \S~\ref{linelist}.

\subsection{Mass-Constrained Retrieval} \label{res:mass-const}
\begin{figure*}[ht!]
\gridline{\fig{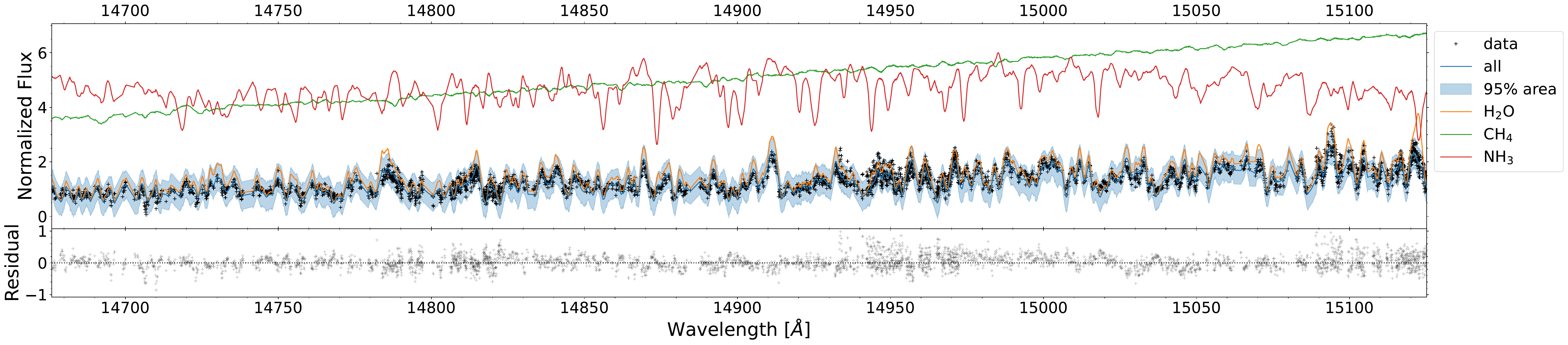}{\textwidth}{(a) Orders 56--58}}
\gridline{\fig{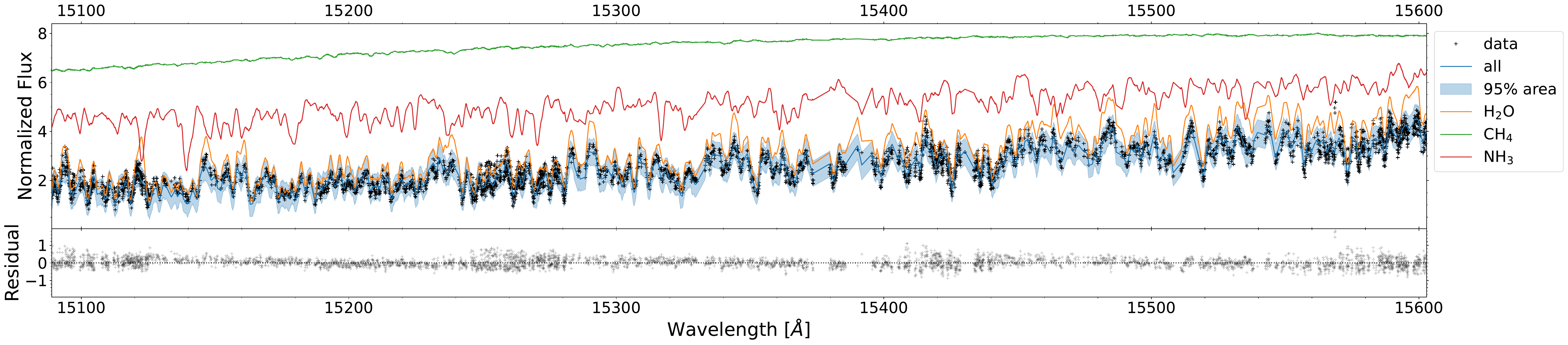}{\textwidth}{(b) Orders 59--61}}
\gridline{\fig{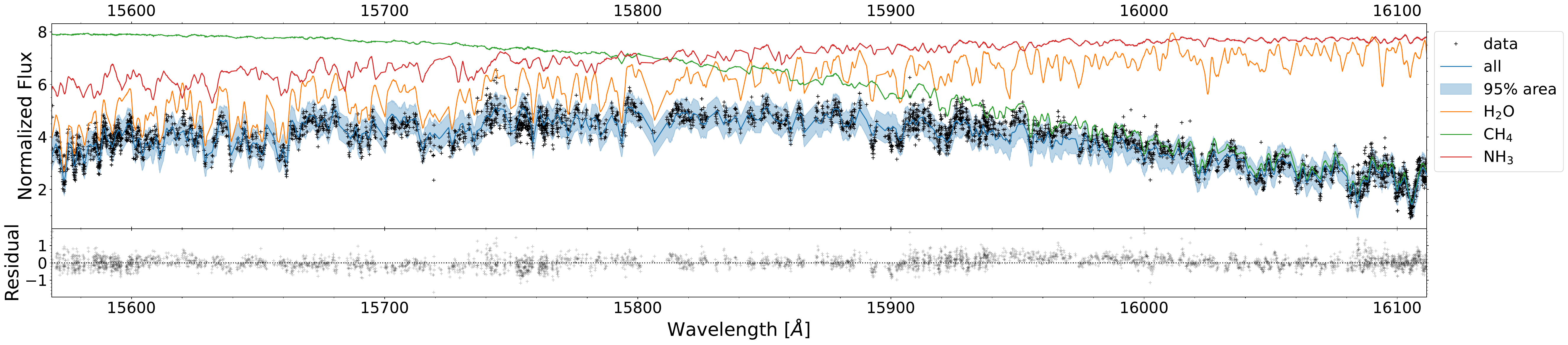}{\textwidth}{(c) Orders 62--64}}
\gridline{\fig{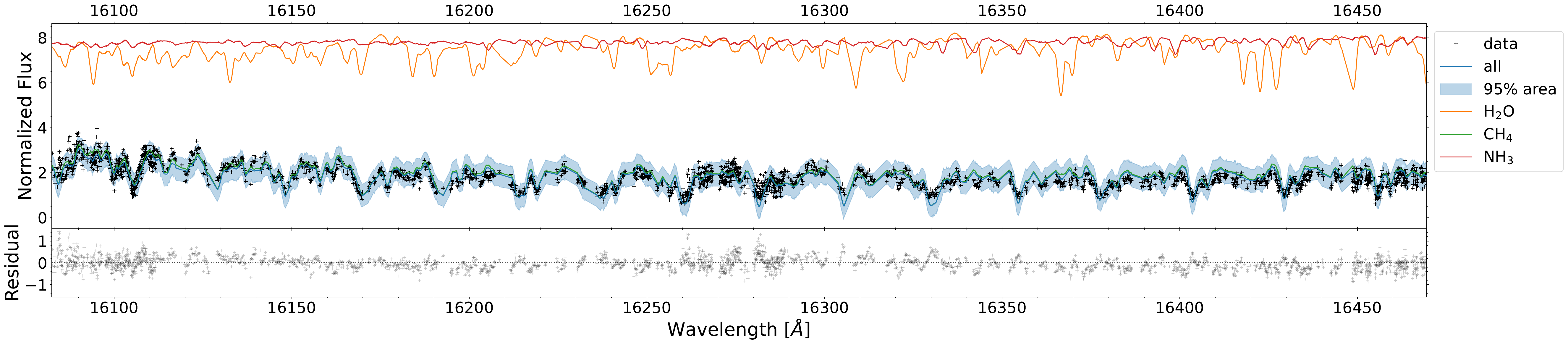}{\textwidth}{(d) Orders 65--66}}
\caption{Result of the mass-constrained retrieval for the spectral orders used. The blue line and shaded region correspond to the median and 95\% credible interval of the spectrum model, respectively. 
The black points indicate our observed Subaru/IRD data.
Also plotted are the median spectrum models calculated with the obtained samples, but only including the specific molecular absorption and CIA to clarify the contribution of each molecule (note that these models are presented without any vertical offset); \mo{H_2O} (orange line), \mo{CH_4} (green line), and \mo{NH_3} (red line). The observed data is normalized by the flux value at the median wavelength of the IRD spectral order 57.
{The bottom panels show the residuals between the observed data and the median model that includes all species (blue line).}
\label{fig:spec}}
\end{figure*}

\begin{figure*}[ht!]
\includegraphics[width=\textwidth]{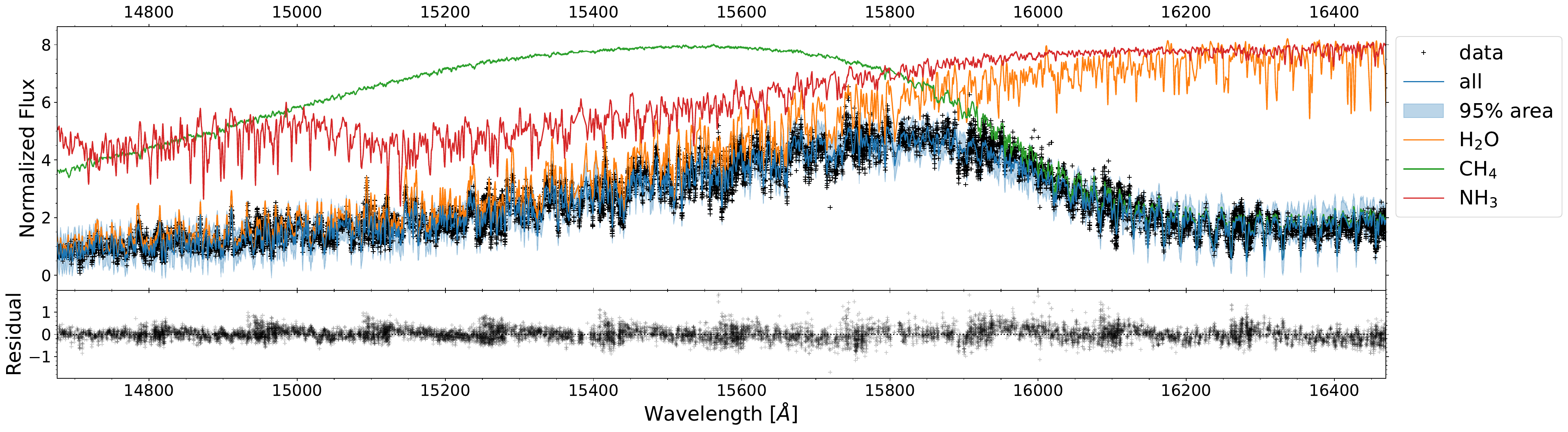}
\caption{{Same as Fig.~\ref{fig:spec}, but showing the wavelength range covered by all the spectral orders analyzed in this study.}
\label{fig:spec_all}}
\end{figure*}

\begin{deluxetable}{lcc}
\tablenum{2}
\tablecaption{Retrieved and derived parameter values of \target \label{tab:values}}
\tablewidth{0pt}
\tablehead{
\colhead{Parameter} & \colhead{Mass-constrained} & \colhead{Mass-free}
}
\startdata
$M$ & {${71.47}^{+0.57}_{-0.61}$} & {${106.71}^{+18.96}_{-16.63}$} \\
$\log_{10} g$ & {${5.01}^{+0.03}_{-0.03}$} & {${5.20}^{+0.09}_{-0.09}$} \\
$\log_{10} x_\mathrm{H_2O}$ & {${-2.85}^{+0.02}_{-0.02}$} & {${-2.73}^{+0.06}_{-0.06}$} \\
$\log_{10} x_\mathrm{CH_4}$ & {${-2.97}^{+0.02}_{-0.02}$} & {${-2.83}^{+0.07}_{-0.06}$} \\
$\log_{10} x_\mathrm{NH_3}$ & {${-4.56}^{+0.03}_{-0.03}$} & {${-4.42}^{+0.07}_{-0.07}$} \\
$T_0$ & {${1000.23}^{+4.49}_{-4.36}$} & {${1000.01}^{+4.10}_{-4.53}$} \\
$\alpha$ & {${0.088}^{+0.001}_{-0.001}$} & {${0.087}^{+0.001}_{-0.001}$} \\
RV & {${5.96}^{+0.13}_{-0.11}$} & {${5.98}^{+0.13}_{-0.13}$} \\
$v \sin{i}$ & {${18.51}^{+0.19}_{-0.31}$} & {${18.13}^{+0.35}_{-0.32}$} \\
$\sigma_\mathrm{j}$ & {${0.265}^{+0.002}_{-0.002}$} & {${0.265}^{+0.002}_{-0.002}$} \\
\hline
$m_H$ & {${14.40}^{+0.07}_{-0.07}$} & {${14.36}^{+0.08}_{-0.06}$} \\
\hline
$R$ & {${1.31}^{+0.05}_{-0.05}$} & {${1.29}^{+0.05}_{-0.04}$} \\
C/O & {${0.77}^{+0.01}_{-0.01}$} & {${0.78}^{+0.01}_{-0.01}$}
\enddata
\tablecomments{The presented errors correspond to the {68\% credible interval} values.
{The middle panel displays the modeled $H$-band magnitude.}
The bottom panel shows the derived values of the radius in units of Jupiter radius $R_\mathrm{J}$ and the C/O ratio from the retrieved parameters.}
\end{deluxetable}

\begin{figure*}[ht!]
\plotone{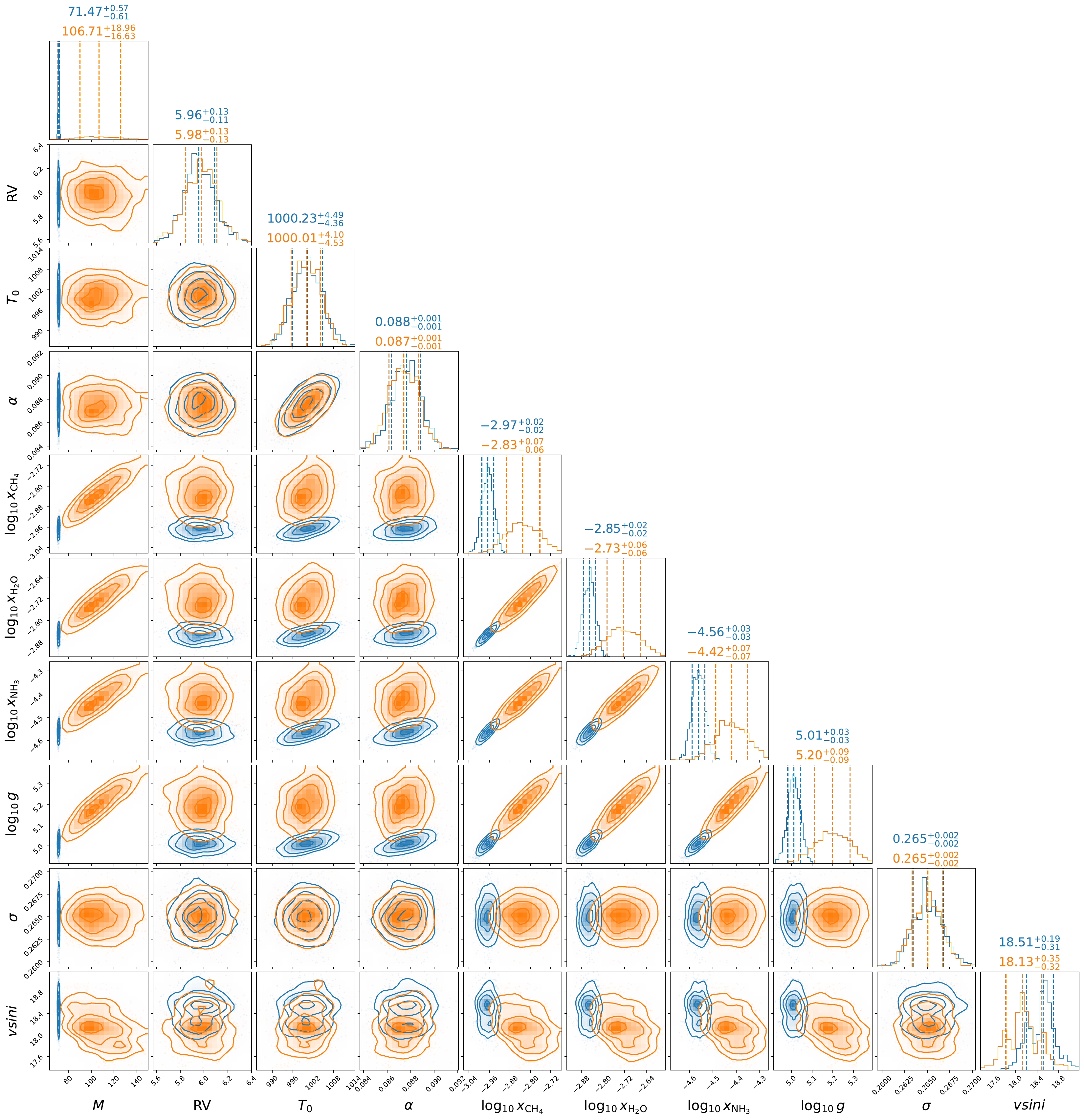}
\caption{Corner plot for the mass-constrained (blue) and mass-free (orange) retrievals.
The vertical dashed lines in the posterior probability distribution plots indicate the median value and the {68\%} credible interval for each parameter.
The units for each parameter are the same as those used in Table~\ref{tab:params}.
\label{fig:corner}}
\end{figure*}

In Figures~\ref{fig:spec} {and \ref{fig:spec_all}}, we show the 
resultant
spectrum fit of the mass-constrained %spectral 
retrieval.
As in previous studies of this object \citep[e.g.,][]{2000ApJ...541..374S, 2022ApJ...940..164C} and observations of other T-type brown dwarfs, the $H$-band spectrum is dominated by absorption by \mo{H_2O} at shorter wavelengths and \mo{CH_4} at longer wavelengths.
This can be clearly seen when comparing the median spectrum (blue line) and those calculated only including the specific molecular contribution and CIA (orange line for \mo{H_2O} and green line for \mo{CH_4}).
%While 
\citet{2000ApJ...541..374S} claimed the tentative detection of \mo{NH_3} based on a medium-resolution spectrum.
We find that \mo{NH_3} indeed contributes to shaping the $H$-band spectrum{, albeit less significantly than \mo{H_2O} and \mo{CH_4}}.
%, which is demonstrated by the difference of the median spectrum calculated including (blue line) and excluding (red line) the \mo{NH_3} contribution\footnote{
The need for \mo{NH_3} can also be understood from its Gaussian-like posterior probability distribution presented in a corner plot (blue in Figure~\ref{fig:corner}).
The values of the parameters we retrieved are summarized in Table~\ref{tab:values}.
%Also, Figure~\ref{fig:corner} presents a corner plot for the mass-constrained retrieval (blue) along with that for the mass-free retrieval (orange), which we will present in \S~\ref{res:mass-free}.
Below, we discuss the obtained values from the mass-constrained retrieval.

\subsubsection{Chemistry {and thermal structure}}
\begin{figure*}[ht!]
\includegraphics[width=\textwidth]{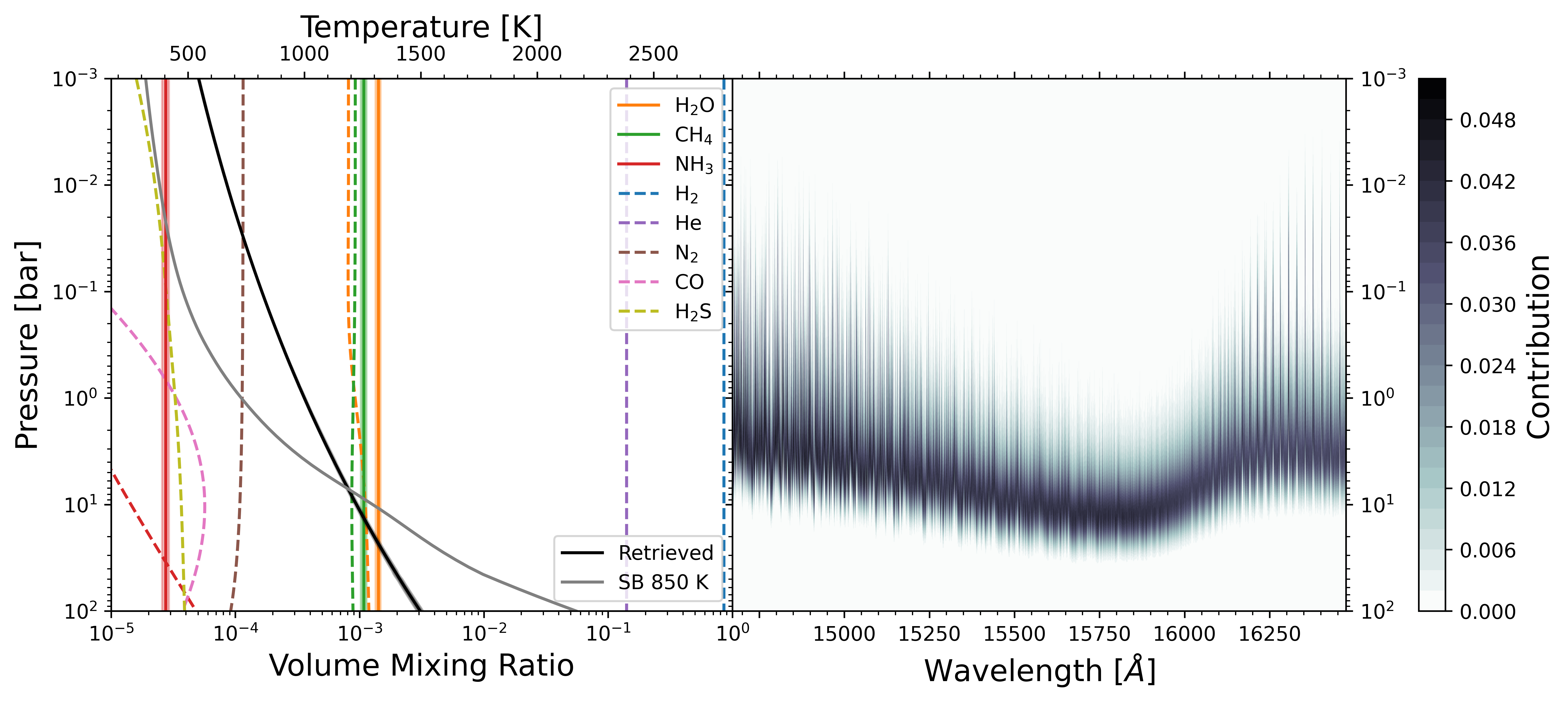}
\caption{(Left) Retrieved abundances of \mo{H_2O} (orange), \mo{CH_4} (green), and \mo{NH_3} (red) for the mass-constrained retrieval plotted along with the temperature--pressure profile ({black}). The solid lines show the median values, while the shaded regions indicate the {68\%} %1$\sigma$ %confidence 
credible interval.
For reference, the thermochemical equilibrium abundances calculated with {\tt Fastchem} are shown by dashed lines.
{The Sonora Bobcat temperature--pressure profile for $T_\mathrm{eff} = 850$~K, $\log_{10} g = 5.0$, and solar metallicity and C/O ratio \citep{2021ApJ...920...85M} is also shown as a gray line.}
(Right) Contribution function for the spectrum model calculated with the median values of the parameters of the mass-constrained retrieval.
\label{fig:chem}}
\end{figure*}

To interpret the retrieved abundances, we used {\tt FastChem}\footnote{\url{https://github.com/NewStrangeWorlds/FastChem}} \citep{2018MNRAS.479..865S} to calculate their thermochemical equilibrium abundances.
%(dotted lines in Fig.~\ref{fig:chem}a).
In that calculation, we simply adopted the median values of $T_0$ and $\alpha$ presented in Table~\ref{tab:values} for the temperature--pressure profile, owing to their tiny bounds.
For the elemental abundances of carbon and oxygen, we used the measured values for the host star, $\log \epsilon_{C} = 8.73$ and $\log \epsilon_{O} = 8.90$ \citep{2015AJ....150...53N} with the definition of $\log \epsilon_{H} = 12.00$.
For the other elements, we adopted a metallicity of $0.30$~dex from the overall metallicity measurement of the host star in \S~\ref{sec:host}. We used the solar elemental abundances of \citet{2021A&A...653A.141A} and ignored the effect of condensation for simplicity.

The left panel of Figure~\ref{fig:chem} shows the comparison between the retrieved abundances of \mo{H_2O} (orange line), \mo{CH_4} (green line), and \mo{NH_3} (red line), and their thermochemical equilibrium abundances (corresponding dashed lines), along with the retrieved temperature--pressure profile ({black} line).
%It can be seen that 
The retrieved abundances of \mo{H_2O} and \mo{CH_4} are both consistent with their thermochemical equilibrium abundances calculated with the elemental abundances of the host star.
%Assuming that all the carbon and oxygen is present in the form of \mo{H_2O} and \mo{CH_4}, respectively, we can derive the C/O ratio as ${0.76}^{+0.01}_{-0.01}$, which is, of course, also consistent with that of the host star
%\citep[$0.68 \pm 0.12$;][]{2015AJ....150...53N}.
On the other hand, the abundance of \mo{NH_3} is larger than the thermochemical equilibrium abundance in the photosphere at $\sim 1-10$~bar (see the right panel of Fig.~\ref{fig:chem}), possibly indicating a quenching process due to vertical mixing.
A retrieval analysis, including the chemically-based quenching process \citep[e.g.,][]{2021A&A...656A..90K}, is beyond the scope of this study.

Assuming that all the carbon and oxygen is present in the form of \mo{H_2O} and \mo{CH_4}, respectively, we can derive the C/O ratio as ${0.77}^{+0.01}_{-0.01}$, which is
%, of course, 
also consistent with that of the host star
\citep[$0.68 \pm 0.12$;][]{2015AJ....150...53N}.

{For reference, the Sonora Bobcat model\footnote{\url{https://zenodo.org/records/5063476}} for $T_\mathrm{eff} = 850$~K, $\log_{10} g = 5.0$, and solar metallicity and C/O ratio \citep{2021ApJ...920...85M} is shown as a gray line. In contrast, our retrieved temperature--pressure profile (black line) is significantly steeper.
This discrepancy likely arises from our adoption of a simple power-law parameterization, which cannot reproduce the more complex structure of the Sonora Bobcat profile. Nevertheless, we consider the retrieved temperature gradient near the photosphere ($\sim 1-10$~bar) to be reliable, owing to the high-sensitivity of high-resolution spectroscopy to temperature via numerous line-strength ratios.
Implementing a more flexible thermal-profile parameterization will be explored in future work.}

\subsubsection{Inferred mass--radius and binary scenario}
In this retrieval, we used the constraint on the $H$-band magnitude,
%information from the photometric observation, 
which relates to the luminosity $L \propto R^2 T^4$.
Here, $R$ and $T$ are the radius and temperature.
Since the high-resolution spectrum is highly sensitive to temperature through the numerous line strength ratios, we can infer the radius by breaking the degeneracy between the radius and temperature, and compare the inferred mass--radius values
%relation 
with the evolutionary model prediction.

\begin{figure}[ht!]
\plotone{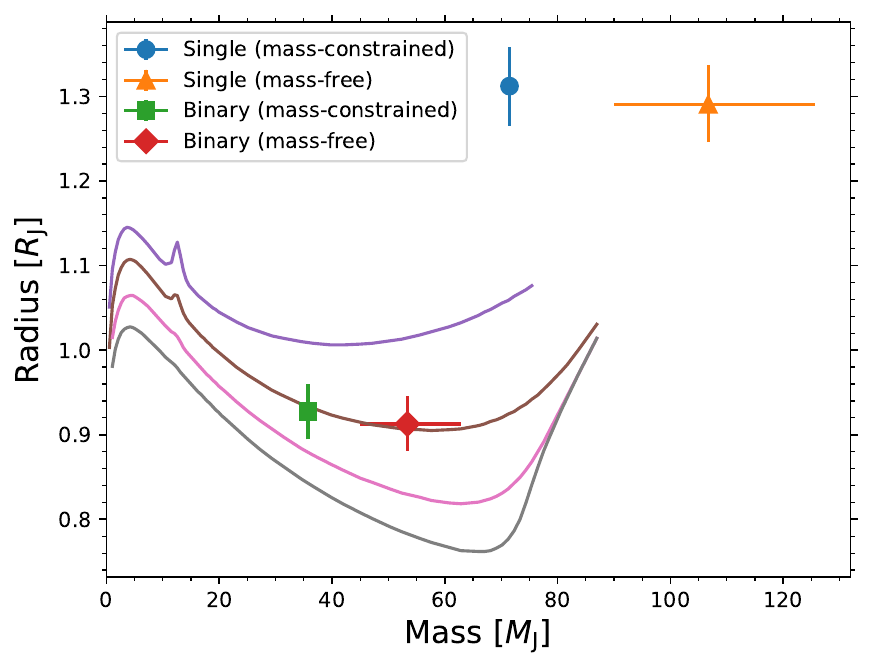}
\caption{Derived masses and radii for the mass-constrained (blue circle) and mass-free (orange triangle) retrievals.
The Sonora Bobcat evolutionary models from \citet{2021ApJ...920...85M} (solar metallicity and C/O ratio) for the ages of 0.4 (purple line), 1.0 (brown line), 3.0 (pink line), and 10.0~Gyr (gray line) are also plotted.
For reference, the corrected mass--radius values
%relations 
for the equal-mass binary scenario are also presented: green square and red diamond points for the mass-constrained and mass-free retrievals, respectively.
\label{fig:M-R}}
\end{figure}

Based on the retrieved mass\footnote{Note that we adopted a narrow prior probability distribution of $\mathcal{N}_t (71.4, 0.6, \mathrm{low = 1.0})$ for the mass.} and gravity, we derived the radius of the object as {${1.31}^{+0.05}_{-0.05}$}.
We compared the obtained mass--radius values
%relation 
with the Sonora Bobcat evolutionary models\footnote{\url{https://zenodo.org/records/5063476} {\citep{marley_2021_5063476}}} \citep[][the case of solar metallicity and C/O ratio]{2021ApJ...920...85M} in Figure~\ref{fig:M-R}.
While the ages of M dwarfs are hard to determine, \citet{2020AJ....160..196B} tentatively estimated the age of \host as $2.6 \pm 0.5$~Gyr based on its activity.
Comparing the prediction from the evolutionary models for such ages, namely 1.0 (brown line) and 3.0 (pink line) Gyr cases, it can be seen that the derived radius (blue circle in Fig.~\ref{fig:M-R}) is exceptionally large.
Thus, to account for the required large surface area, we discuss the binary scenario in the following paragraph.
%based on the results we obtained.

{We} note that a similar discrepancy was pointed out by \citet{2021AJ....162..301B}, who also attributed it to an unresolved binary scenario.
{More recently,
\citet{2024Natur.634.1070X}\footnote{As noted in the Introduction, this paper was published while the present manuscript was under review.} resolved \target into two components using the GRAVITY interferometer and CRIRES+ spectrograph at the Very Large Telescope, reporting individual masses of $38.1 \pm 1.0$ and $34.4 \pm 1.5$~$M_\mathrm{J}$.
%Thus, assuming \target is an equal-mass binary provides a natural explanation for the observed discrepancy.
}

{Indeed}, the observed discrepancy can be well explained if we assume that \target is composed of an equal-mass binary.
In the equal-mass binary case, since they would follow the same thermal evolution thus have exactly the same effective temperature, the spectrum of the two bodies would be exactly the same.
This is consistent with the excellent fit by a single-body assumption (Figs.~\ref{fig:spec} {and \ref{fig:spec_all}}).
Also, the low luminosity for the measured mass reported by \citet{2021AJ....162..301B} would be resolved due to the efficient cooling thanks to the increased surface area.
%Given that the observed spectrum is well-fitted by a single-body spectrum (Fig.~\ref{fig:spec}), we hypothesize that even if \target is indeed an unresolved binary, it is close to a twin binary.
On the other hand, 
%Also, 
if \target is a binary with a large mass ratio, it would make little difference in its thermal evolution.
Thus, for simplicity, we consider the case where the object is composed of two objects with equal mass and the same effective temperature, and examine how the interpretation of the obtained retrieval result would change.

The gravity of the single-body case $g_\mathrm{s}$ is expressed as $g_\mathrm{s} = G M_\mathrm{s}/R_\mathrm{s}^2$, where $G$ is the gravitational constant, and $M_\mathrm{s}$ and $R_\mathrm{s}$ are the mass and radius for the single-body case, respectively.
To conserve the observed luminosity, the radius for the binary case needs to be $R_\mathrm{b} = R_\mathrm{s}/\sqrt{2}$.
Considering that the mass for the equal-mass binary case needs to be $M_\mathrm{b} = M_\mathrm{s}/2$, the gravity, which is a parameter for the spectral calculation, is the same as in the single case, namely $g_\mathrm{b} = g_\mathrm{s}$.
Thus, we can use the same retrieval result for the equal-mass binary case by interpreting $M_\mathrm{b} = M_\mathrm{s}/2$ and $R_\mathrm{b} = R_\mathrm{s}/\sqrt{2}$.

The mass--radius point
%relation 
in this equal-mass binary case is plotted as the green square in Fig.~\ref{fig:M-R}.
While this is a simple argument, the mass--radius values
%relation 
for the equal-mass binary case exhibit those much more consistent with the evolutionary model prediction.
This indicates that the binary scenario is supported not only by photometric and astrometric observations \citep{2021AJ....162..301B}, {and interferometry and radial velocity measurements \citep{2024ApJ...974L..30W,  2024Natur.634.1070X}}, but also by high-resolution spectroscopy \citep[][and this study]{2024Natur.634.1070X}, thanks to its excellent sensitivity to temperature {in the case of this study}.
%Future radial velocity and interferometry observations are desired to investigate this binary scenario.

\subsection{Mass-Free Retrieval}  \label{res:mass-free}
For %widely separated 
long-period
planetary-mass objects and abundant isolated brown dwarfs, dynamical mass measurements by radial velocity and astrometric observations are challenging or impossible, requiring mass estimations that rely on evolutionary models despite the generally high uncertainty in their age and metallicity.
%Also, 
Recent advancements in dynamical mass measurements for a limited number of brown dwarfs orbiting host stars have 
%allowed a comparison between the measured dynamical mass and that estimated based on the evolutionary models, namely photometric mass, revealing their discrepancy with a possible correlation with gravity 
revealed a discrepancy with the masses based on the evolutionary models that may be correlated with gravity \citep{2021AJ....162..301B}.
In this context, it is essential to develop another mass inference method for long-period
%wide-orbit 
planetary-mass objects and isolated brown dwarfs, such as a spectroscopic method.
%, is %essentially 
%needed.

We explore the possibility of inferring gravity from high-resolution spectra as, in that case, the mass can be derived by the information on gravity and radius.
With the assumption of hydrostatic equilibrium, the contribution to the optical depth by the line absorption of molecules $i$, $\tau_{\mathrm{line}, i}$, is written as
\begin{eqnarray} \label{eq:tau_line}
    d \tau_{\mathrm{line}, i} &=& - \frac{x_i \sigma_i}{\mu m_u g} dP \\
    &=& - \frac{y_i \sigma_i}{M_i m_u g} dP,
\end{eqnarray}
where $g$ is the gravity and $\mu$ and $m_u$ are the mean molecular weight and atomic mass unit, respectively.
$x_i$, $y_i$, $\sigma_i$, and $M_i$ are the volume mixing ratio, mass mixing ratio, cross-section, and molecular mass of the species $i$, respectively.
Noticeably, the degeneracy between the gravity and mass mixing ratio is inherent in the optical depth by molecular line absorption.
On the other hand, the contribution to the optical depth of the collision-induced absorption (CIA) between molecules $i$ and $j$ with their number densities of $n_i$ and $n_j$, $\tau_{\mathrm{CIA}, i, j}$, is written as
\begin{eqnarray}
    d \tau_{\mathrm{CIA}, i, j} &=& - \beta_{i, j} n_i n_j \frac{k_\mathrm{B} T}{\mu m_u g} \frac{dP}{P},
\end{eqnarray}
where $\beta_{i, j}$ is the CIA absorption coefficient and $k_\mathrm{B}$ is the Boltzmann constant.
Considering that for hydrogen-dominated atmospheres of gas giant planets and brown dwarfs, $\mathrm{H_2}$--$\mathrm{H_2}$ CIA dominates the spectral continuum in a specific wavelength range and the possible range of $n_\mathrm{H_2}$ is relatively small, we can infer the gravity and thus extract the mass information from the spectrum in the CIA wavelength region.
Indeed, as demonstrated in Figure~\ref{fig:CIA} in Appendix~\ref{sec:CIA}, we have confirmed that while the mass mixing ratio and gravity are degenerate in the optical depth by molecular line absorption, the CIA optical depth is irrelevant to this degeneracy.
Thus, the high-resolution spectrum offers another method of mass estimation also applicable to long-period
%wide-orbit 
planetary-mass objects and isolated brown dwarfs.

%In this subsection, we present the result of the retrieval, adopting the uniform prior of $1$--$150$~$M_\mathrm{J}$ for the mass.
%As shown in Table~\ref{tab:values} and Figure~\ref{fig:corner}, 
The results of the mass-free retrieval are shown in Table~\ref{tab:values} and Fig.~\ref{fig:corner} {(orange)}.
% The values of the parameters 
% %we retrieved for the mass-constrained retrieval 
% are similar to those we retrieved for the mass-{constrained} retrieval, except for the mass, which has, of course, much larger uncertainty.
{In the mass-free case, the retrieved gravity, and consequently the inferred mass, is higher; we discuss possible reasons for this in the following paragraph.
Although the CIA wavelength region helps to partially break the degeneracy between gravity and molecular mixing ratios in the optical depth of molecular line absorption (see Eq.~\ref{eq:tau_line}), a residual correlation remains (Fig.~\ref{fig:corner}). This leads to higher retrieved abundances due to the elevated gravity compared to the mass-constrained case.
Consequently, the uncertainties on gravity and molecular abundances are larger in the mass-free retrieval, demonstrating that an independent mass constraint significantly improves our ability to determine these parameters.
%Comparing the results of the two retrievals in detail, it can be seen that 
%the constraint on the mass helps determine the molecular abundances and gravity.
In contrast, 
%while
the parameters 
%for the temperature-pressure profile, 
$T_0$, $\alpha$, RV, and $v \sin{i}$ are essentially unaffected by the mass prior: their retrieved values and uncertainties are nearly identical between the mass-free and mass-constrained retrievals, as these quantities are encoded primarily in the line profiles.}
%, probably because the high-resolution spectrum is highly sensitive to temperature through the line strength ratios.
% massをconstraintするとabundanceの制限が強まる
% except for mass
% high-resoのlineから得ているパラメータはあまり変わらないけど、gとかvmrとかはかわる

\begin{figure}[ht!]
\plotone{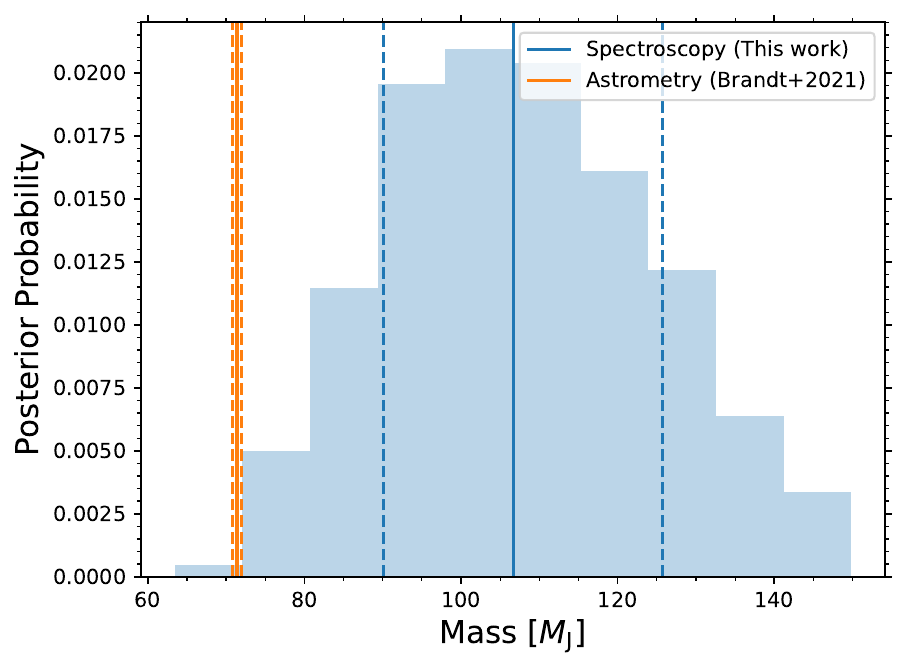}
\caption{Posterior probability distribution of the mass compared to the dynamical mass measurement (orange line) from \citet{2021AJ....162..301B}.
The vertical solid and dashed lines indicate the median value and the {68\%} credible interval, respectively.
\label{fig:mass}}
\end{figure}

Figure~\ref{fig:mass}
compares the posterior probability distribution of the mass from the mass-free retrieval to the dynamical mass measurement of \citet{2021AJ....162..301B}.
%presents the posterior probability distribution of the mass from the mass-free retrieval, which is compared to the dynamical mass measurement of \citet{2021AJ....162..301B}.
While the retrieved mass differs from the dynamical mass by $\sim 2\sigma$, considering the scattered mass estimate by previous studies (see Figure~6 of \citet{2022ApJ...935..107H} for the summary plot),
our result still demonstrates the ability of high-resolution spectroscopy to constrain the mass.
{It is worth noting that our mass estimate from high-resolution spectroscopy tends to be higher than the reliable dynamical mass measurement of \citet{2021AJ....162..301B}. By contrast, the medium-resolution spectral analysis of \citet{2022ApJ...935..107H} generally yields lower mass estimates.
While a detailed investigation of this systematic difference is beyond the scope of this work, it may be related to the use of different datasets and assumptions about the temperature--pressure profile.}

%Below, we discuss the possible reason for the discrepancy between the spectroscopic and astrometric masses.
%First, %we have employed a relatively simple temperature--pressure profile.
{In addition to the binarity of the target,} one possible reason for the discrepancy between the spectroscopic and astrometric masses is that
the power-law temperature--pressure profile we have employed may be too simplistic.
If the actual temperature in the deeper CIA pressure region is hotter than extrapolated based on the retrieved power-law,
%due to the limitations of the adopted power-law profile, 
it would lead to overestimate
the gravity
%would be higher than it really is
%is by trying 
due to the reproduction of the large flux at the CIA wavelength region.
Indeed, when compared to the temperature--pressure profile of the Sonora Bobcat model \citep{2021ApJ...920...85M} {at the relevant effective temperature}, our retrieved thermal profile in the deep atmosphere is cooler {(see the left panel of Fig.~\ref{fig:chem})}.
Adopting a more flexible thermal profile to account for the different temperature gradient in the deeper atmosphere is a subject of future study.
Second, we have ignored the effect of clouds, but we consider that this is irrelevant to the observed discrepancy.
If clouds are neglected for the spectrum calculation, the simulated atmosphere will be optically thinner, mimicking a lower-gravity condition.
Thus, ignoring the effect of clouds rather works towards underestimating the mass.
Finally, the trend that remained through the data reduction process
%blaze function correction 
could be also related.
%As described in \S~\ref{sec:reduction}, we corrected the blaze function by normalizing the target spectrum by the flat data, namely, assuming the same function for both the target and flat data.
As described in \S~\ref{sec:reduction}, we corrected the temperature dependence of the flat lamp by estimating its black body radiation temperature from the comparison of the spectrum of a telluric standard star with that of a similar spectral type.
Also, the possible difference in the efficiency of the adaptive optics (AO) between the reference (\host) and target produce the wavelength-dependent trend in the amount of light injected into the fiber.
Such an assumption and factor involve uncertainty and thus might yield an artifact wavelength trend in the final spectrum.
Since the difference in the spectrum produced by that of gravity is subtle (Fig.~\ref{fig:CIA}), these points might affect the mass estimation.

% \subsection{Binary retrieval}
% mass ratio of $q = 0.90 \pm 0.04$
% \begin{equation}
%     M_1 = \frac{1}{1 + q} M_\mathrm{tot}
% \end{equation}
% \begin{equation}
%     M_2 = \frac{q}{1 + q} M_\mathrm{tot}
% \end{equation}
% \begin{equation}
%     \sigma_{M_1} = \sigma_{M_2} = \frac{\sqrt{\frac{1}{\left( 1+q \right)^2} M_\mathrm{tot}^2 \sigma_q^2 + \sigma_{M_\mathrm{tot}}^2}}{1 + q}
% \end{equation}

\subsection{Molecular Line List Validation} \label{linelist}
{High signal-to-noise ratio}, high-resolution spectra of brown dwarfs offer a valuable opportunity to test the current molecular line lists at a unique temperature range relevant to exoplanets and brown dwarfs ($\sim 500$--$2500$~K), which is usually hard to replicate for laboratory molecular spectroscopy.
Recently, \citet{2022MNRAS.514.3160T} and \citet{2023ApJ...953..170H} compared the ExoMol YT34to10 \citep{2017A&A...605A..95Y} and HITEMP \citep{2020ApJS..247...55H} methane line lists for the observed spectra of late T-type brown dwarfs they observed and they both found better {match} for the HITEMP.
Recently, ExoMol released an upgraded MM methane line list \citep{2024MNRAS.528.3719Y}.

To examine ExoMol and HITEMP/HITRAN molecular line lists
%current molecular line lists 
by utilizing the high-resolution spectrum of an observed brown dwarf, we performed an optimization to search for the best-fit spectrum model for each IRD spectral order in the $H$-band, using two molecular line list databases, ExoMol and HITEMP/HITRAN.
We minimized $\chi^2$ (Eq.~\ref{eq:scaling}) using the Adam optimization implemented in JAXopt \citep{deepmind2020jax, 2021arXiv210515183B}.
We considered the same parameter sets presented in Table~\ref{tab:params}.
But, we excluded the mass, which we fixed at 71.4~$M_\mathrm{J}$, and included the scaling factor $a$.
For the molecular opacity, in addition to the molecules we considered in the retrieval, namely \mo{H_2O}, \mo{CH_4}, and \mo{NH_3}, we also considered \mo{CO} and \mo{H_2S} to search for their possible absorption signatures.
For the ExoMol case, we tried old and new methane line lists, YT34to10 \citep{2017A&A...605A..95Y} and MM \citep{2024MNRAS.528.3719Y}.
For the HITEMP/HITRAN case, we used the HITEMP database for \mo{H_2O}, \mo{CH_4}, and \mo{CO}, while the HITRAN was used for \mo{NH_3} and \mo{H_2S} due to the absence of these molecules in the HITEMP database.

\begin{figure*}[ht!]
\plotone{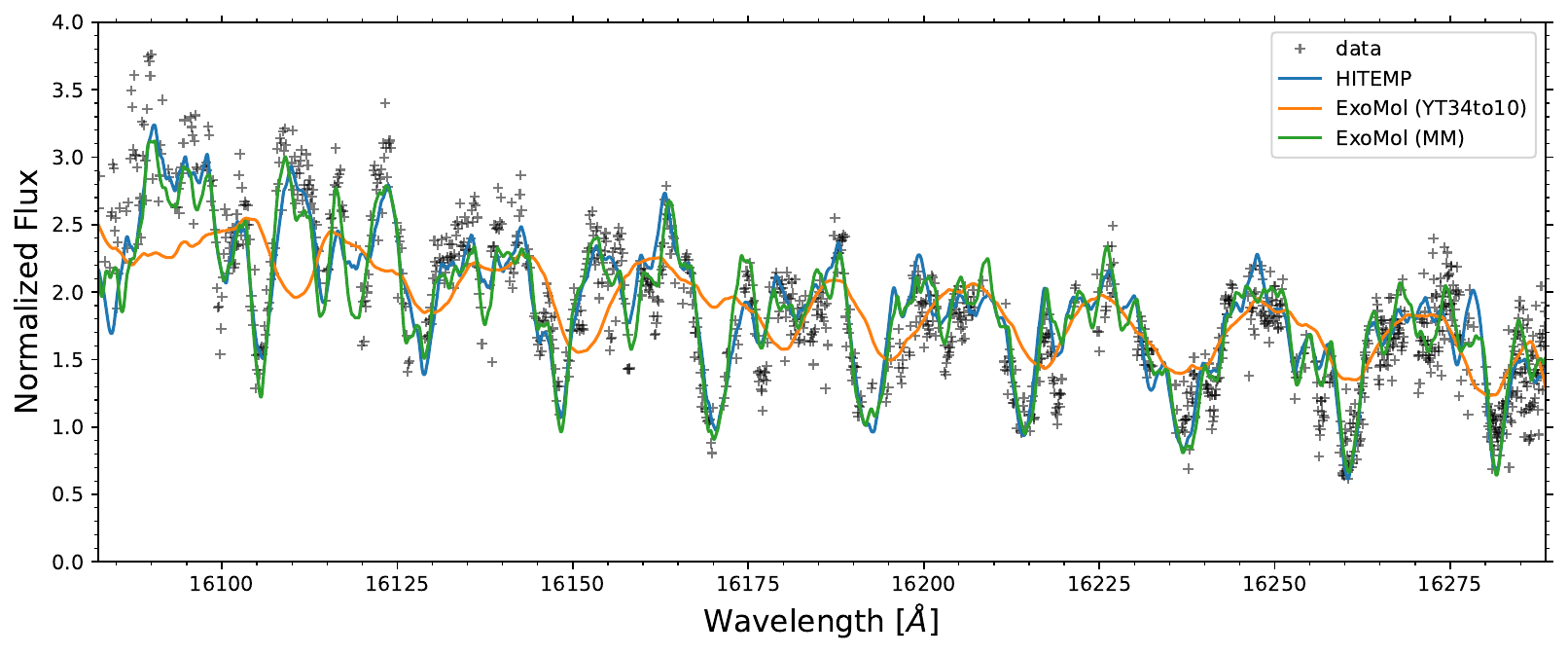}
\caption{Best-fit models calculated with different molecular line list databases for the IRD spectral order 65, where methane absorption dominates the spectrum: HITEMP (blue line), Exomol/YT34to10 (orange line), and ExoMol/MM (green line).
Note that the best-fit model for the ExoMol YT34to10 methane line list case results in a quite different parameter space (especially distinct RV and extremely large $v \sin{i}$) to just account for the overall trend due to the disagreement of the line positions of the methane absorption features with the observation.
\label{fig:methane}}
\end{figure*}

Among the molecules we tested, we found a noticeable difference between the databases for \mo{CH_4}.
Figure~\ref{fig:methane} shows the best-fit spectrum models for the IRD spectral order 65, where methane absorption dominates the spectrum.
%Likewise, 
Similar to \citet{2022MNRAS.514.3160T} and \citet{2023ApJ...953..170H}, we found a poorer match for the observed spectrum for our target with the  ExoMol YT34to10 methane line list \citep{2017A&A...605A..95Y}.
On the other hand, both the HITEMP methane line list \citep{2020ApJS..247...55H} and the latest ExoMol MM methane line list \citep{2024MNRAS.528.3719Y} reproduced the observed absorption features {relatively well. However, some observed lines, such as those at 16100, 16175, 16220, 16240, and 16255~$\AA$, remain discrepant with all model spectra.}.

\section{Discussion} \label{sec:discussion}
\subsection{Comparison with the Previous Medium-Resolution Spectral Retrieval}
While our retrieved value for the C/O ratio is ${0.77}^{+0.01}_{-0.01}$, similar to that for the host star \citep[$0.68 \pm 0.12$;][]{2015AJ....150...53N}, recent retrieval studies based on a medium-resolution spectrum of \target by \citet{2022ApJ...935..107H} and \citet{2022ApJ...940..164C} both reported a high C/O ratio of $\gtrsim$ 1.0.
{Note that although those studies included \mo{CO} and \mo{CO_2} in their analyses, their retrieved abundances for these species are over an order of magnitude lower than our \mo{H_2O} and \mo{CH_4} values.
Therefore, omitting \mo{CO} and \mo{CO_2} here has only a minor impact on our derived C/O ratio.}
Considering the formation process of such a massive object, namely gravitational instability, however, it is reasonable for the object to have a value similar to that of its host star \citep{2011ApJ...743L..16O}.
Thus, \citet{2022ApJ...940..164C} partially attributed the high C/O ratio they retrieved to oxygen depletion due to cloud formation, although it proved
%was proven 
insufficient to account for the deviation.

\begin{figure*}[ht!]
\plotone{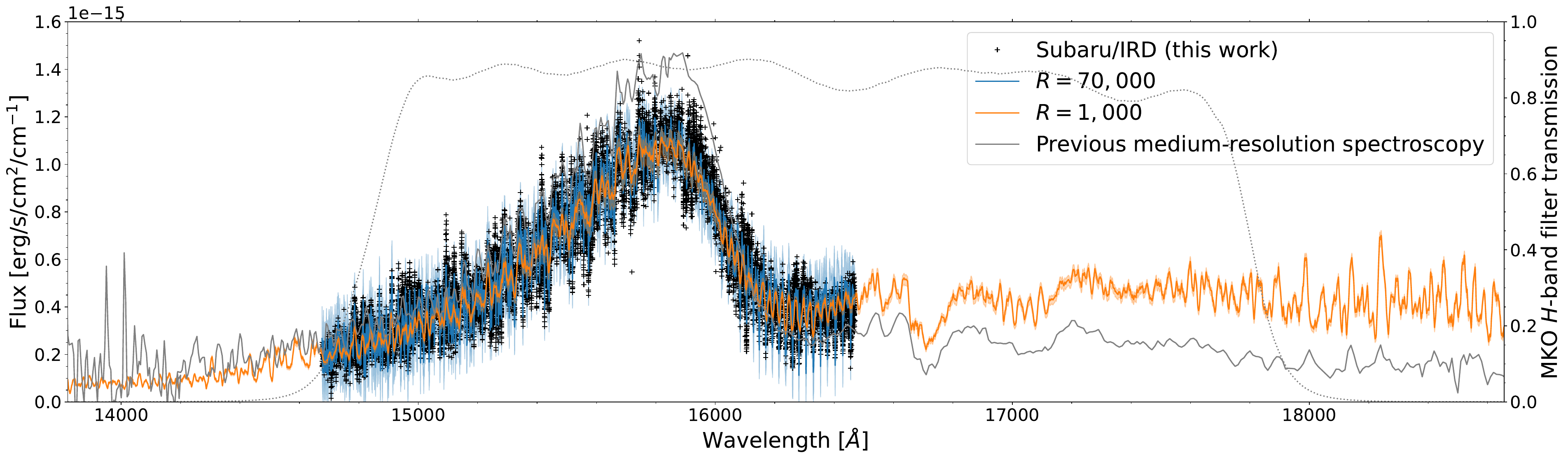}
\caption{Comparison between our Subaru/IRD data (black points) and the compilation of the previously measured medium-resolution spectra \citep[gray line;][]{1996ApJ...467L.101G, 1998ApJ...492L.181S, 1998ApJ...502..932O, 1997ApJ...489L..87N}.
{The median model spectra at $R = 70,000$ (blue line; same as the blue lines in Figs.~\ref{fig:spec} and \ref{fig:spec_all}) and $R = 1,000$ (orange line), computed from our retrieved posterior samples, are shown along with their corresponding 95\% intervals.
The medium-resolution spectrum in this wavelength region \citep{1996ApJ...467L.101G} has a resolution of $R \sim 780 \times \lambda$ below, and $R \sim 390 \times \lambda$ above, 1.585~$\mu$m, where $\lambda$ is the wavelength in $\mu$m.
For reference, we also show the transmission curve of the MKO NSFCam {$H$-band} filter, obtained from the SVO Filter Profile Service \citep{2012ivoa.rept.1015R, 2020sea..confE.182R}, as a gray dotted line aligned with the left vertical axis.}
%For reference, the median and 95\% credible interval of the spectrum model, which are the same as those presented in Fig.~\ref{fig:spec}, are plotted with the blue line and shaded region, respectively. 
%Note that the absolute flux values for our Subaru/IRD data were derived by considering the median sample values of the scaling parameter $a$ in Eq.~(\ref{eq:scaling}).
\label{fig:comp_mid}}
\end{figure*}

In Figure~\ref{fig:comp_mid}, we compare our Subaru/IRD high-resolution spectrum with the compilation of the previously measured medium-resolution spectra \citep{1996ApJ...467L.101G, 1998ApJ...492L.181S, 1998ApJ...502..932O, 1997ApJ...489L..87N}\footnote{\url{http://staff.gemini.edu/~sleggett/LTdata.html}}, %with the flux calibration of \citet{1998AJ....115.2579G}, \citet{1999ApJ...517L.139L}, and \citet{2004AJ....127.3516G}
which \citet{2022ApJ...935..107H} and \citet{2022ApJ...940..164C} used for their retrieval.
{While we adopt a value of $H_\mathrm{MKO} = 14.36 \pm 0.05$~mag \citep{2012ApJ...752...56F}, a joint fit with the normalized high-resolution IRD spectrum (see Eq.~\ref{eq:likelihood}) yields a modeled value of $H_\mathrm{MKO} = 14.40^{+0.07}_{-0.07}$~mag (see Table~\ref{tab:values}).
The $H$-band medium-resolution spectrum was originally calibrated to $H_\mathrm{UKIRT} = 14.35 \pm 0.05$~mag \citep{1999ApJ...517L.139L}.
Therefore, in Fig.~\ref{fig:comp_mid}, we plot the medium-resolution spectrum scaled to match $H_\mathrm{MKO} = 14.40$~mag.}

As can be seen, the overall trend is different between the two observed spectra.
%, in addition to the difference in the radiative transfer models used, 
It is difficult to ascertain the reason for this discrepancy: {it could stem from wavelength-dependent systematics in either in our reduction or the medium-resolution data, or from intrinsic variability in the atmosphere of \target. To address this,}
it is important to increase the sample of targets for which both broadband medium-resolution spectroscopy and narrowband high-resolution spectroscopy are performed.
This is essential to pave the way for comprehensive atmospheric characterization since high-resolution and medium/low-resolution spectroscopy are sensitive to different parameters.

\subsection{Temperature--Pressure Profile}
In this study, we have adopted a simple power-law temperature--pressure profile.
Given the relatively narrow wavelength region, namely the limited pressure region, we analyzed (see the right panel of Fig.~\ref{fig:chem}), we expect that the effect of this assumption is relatively small.
However, as discussed in \S~\ref{res:mass-free}, the discrepancy between the spectroscopic mass we retrieved and the astrometric mass of \citet{2021AJ....162..301B} could arise from this simple assumption. We leave adopting a more flexible thermal profile for future studies.

% by H.K. 7/22 edited by Y.K. 8/26
% \subsection{Importance of Companion Brown Dwarfs as Benchmark Objects}
% In this paper, we have demonstrated the usefulness of a companion brown dwarf as a benchmark object for atmospheric characterization. This usefulness lies in the ability to compare elemental abundances, such as the C/O ratio, with the primary star and to verify/calibrate the spectroscopic mass using the dynamical mass. Additionally, from an observational standpoint, the primary star, being a normal star, serves as a natural guide star, allowing the use of adaptive optics (AO). This provides the significant advantage of allowing efficient spectroscopy on faint companions, extending to relatively cool T-type brown dwarfs like \target. Such systems, which are combinations of a normal star and a T-type dwarf, include objects like WISE J072003.20-084651.2 or Scholz's star \citep{2014A&A...561A.113S}. Advancing high-resolution observations of these systems in the future will be beneficial for enhancing the robustness of retrieval techniques and spectroscopic mass estimation.

%current/future inst developments, REACH/HiRISE/KPIC
% about Roman?

\section{Summary} \label{sec:summary}
In this paper, we presented the results of the  high-resolution spectroscopic observations of the archetype T-type brown dwarf \target.
We obtained its near-infrared high-resolution ($R \sim 70,000$) spectrum by the InfraRed Doppler instrument \citep[IRD;][]{2012SPIE.8446E..1TT, 2018SPIE10702E..11K} mounted on the Subaru telescope.
We applied our recently developed high-resolution spectrum code for exoplanets and brown dwarfs \ej \citep{2022ApJS..258...31K, exojax2} to the IRD spectrum reduced by PyIRD \citep{pyIRD} to perform spectral retrieval.

%As a result, u
Unlike previous studies of the medium-resolution spectral retrieval for \target, we obtained the abundances of \mo{H_2O} and \mo{CH_4} and thus the C/O ratio consistent with the abundances of the host star, which is compatible with the formation process for such a massive object.
On the other hand, the retrieved abundance of \mo{NH_3} was larger than the thermochemical equilibrium value calculated from the metallicity of the host star, which might imply an ongoing effect of vertical mixing.

Also, we have found that the retrieved radius is significantly discrepant with evolutionary model prediction.
This can be resolved if we assume that \target is an unresolved equal-mass binary, as \citet{2021AJ....162..301B} proposed {and recently confirmed by \citet{2024ApJ...974L..30W} and \citet{2024Natur.634.1070X}}.
% We have also explored the scenario proposed by \citet{2021AJ....162..301B} that \target is an unresolved binary.
% By leveraging the fact that the gravity for the equal-mass binary case gives exactly the same results as for the single body case, we attempted to interpret our retrieval results for the equal-mass binary case.
% While the retrieved radius for the single-body case strongly deviates from the evolutionary model prediction, that for the equal-mass binary case matches the prediction, indicating that our high-resolution spectroscopic observation also supports the binary scenario.

We further explored the possibility of inferring the mass of the object from the high-resolution spectrum.
While the mass we retrieved from the embedded information of CIA differs from the astrometric mass by $\sim 2\sigma$, given the scattered prediction by previous studies, we consider that this still demonstrates the ability to infer the mass from the high-resolution spectra.
This is especially useful for abundant single brown dwarfs and planetary-mass objects %in wide orbits
with long periods whose mass measurement through radial velocity and astrometric observations is impossible or challenging.

Finally, using the {high signal-to-noise ratio}, high-resolution spectrum we obtained for this relatively cool T-type brown dwarf, we examined the current molecular line lists.
From a comparison of the best-fit model spectra for different molecular line list databases,
we found that the observed spectrum % less matches 
relatively poorly matches the ExoMol YT34to10 methane line list \citep{2017A&A...605A..95Y}, while both the HITEMP methane line list \citep{2020ApJS..247...55H} and the latest ExoMol MM methane line list \citep{2024MNRAS.528.3719Y} reproduce the observed absorption features, {although some observed lines remain discrepant}.

% This study has demonstrated that the observing companion brown dwarfs, such as \target, is useful in various aspects.
% Knowledge of the elemental abundances of the host star, possible dynamical mass measurements through astrometry and radial velocity, and the availability of the primary star as a natural guide star for the use of adaptive optics (AO) not only significantly contribute to realizing an in-depth understanding of the atmosphere and formation/evolution processes of the targets themselves, but also pave the way for current/future characterization of low-mass objects, including brown dwarfs and exoplanets.

% by H.K. 7/22 edited by Y.K. 8/26
In this paper, we have demonstrated the usefulness of a companion brown dwarf as a benchmark object for atmospheric characterization. This usefulness lies in the ability to compare elemental abundances, such as the C/O ratio, with the primary star and to verify/calibrate the spectroscopic mass using the dynamical mass measured through astrometry and radial velocity. Additionally, from an observational standpoint, the primary star, being a normal star, serves as a natural guide star, allowing the use of adaptive optics (AO). This provides the significant advantage of allowing efficient spectroscopy on faint companions, extending to relatively cool T-type brown dwarfs like \target. Such systems, which are combinations of a normal star and a T-type dwarf, include objects like WISE J072003.20-084651.2 or Scholz's star \citep{2014A&A...561A.113S}. Advancing high-resolution observations of these systems in the future will be beneficial for enhancing the robustness of retrieval techniques and spectroscopic mass estimation, and paving the way for current/future characterization of low-mass objects, including brown dwarfs and exoplanets.

%% IMPORTANT! The old "\acknowledgment" command has be depreciated. It was
%% not robust enough to handle our new dual anonymous review requirements and
%% thus been replaced with the acknowledgment environment. If you try to 
%% compile with \acknowledgment you will get an error print to the screen
%% and in the compiled pdf.
\begin{acknowledgments}
We would like to express our sincere appreciation to the referee for their careful reading of our manuscript and many constructive comments, which have been a great help in improving the manuscript.
Y.K. acknowledges support from JSPS KAKENHI Grant Numbers 21K13984, 21H04998, 22H05150, and 23H01224, and the Special Postdoctoral Researcher Program at RIKEN. This study was also supported by JSPS KAKENHI nos. 23H00133 (H.K.). 
Y. Kasagi is supported by JSPS KAKENHI grant No. 24K22912.
M. K. is supported by the JSPS KAKENHI grant No. 24K07108.
S.K.N is supported by JSPS KAKENHI grant No. 22K14092.
HH is supported by JSPS KAKENHI grant No. 21K03653.
M.T. is supported by JSPS KAKENHI grant No.24H00242. 
Numerical computations were in part carried out on GPU and PC clusters at the Center for Computational Astrophysics, National Astronomical Observatory of Japan.
This research has made use of the SVO Filter Profile Service ``Carlos Rodrigo'', funded by MCIN/AEI/10.13039/501100011033/ through grant PID2020-112949GB-I00.
\end{acknowledgments}

%% To help institutions obtain information on the effectiveness of their 
%% telescopes the AAS Journals has created a group of keywords for telescope 
%% facilities.
%
%% Following the acknowledgments section, use the following syntax and the
%% \facility{} or \facilities{} macros to list the keywords of facilities used 
%% in the research for the paper.  Each keyword is check against the master 
%% list during copy editing.  Individual instruments can be provided in 
%% parentheses, after the keyword, but they are not verified.

\vspace{5mm}
\facilities{Subaru (IRD)}

%% Similar to \facility{}, there is the optional \software command to allow 
%% authors a place to specify which programs were used during the creation of 
%% the manuscript. Authors should list each code and include either a
%% citation or url to the code inside ()s when available.

\software{Astropy \citep{2013A&A...558A..33A, 2018AJ....156..123A, 2022ApJ...935..167A},
          corner.py \citep{2016JOSS....1...24F},
          ExoJAX \citep{2022ApJS..258...31K, exojax2},
          JAX \citep{jax2018github},
          NumPy \citep{harris2020array},
          NumPyro \citep{2019arXiv191211554P, JMLR:v20:18-403},
          Matplotlib \citep{Hunter:2007}
          }

%% Appendix material should be preceded with a single \appendix command.
%% There should be a \section command for each appendix. Mark appendix
%% subsections with the same markup you use in the main body of the paper.

%% Each Appendix (indicated with \section) will be lettered A, B, C, etc.
%% The equation counter will reset when it encounters the \appendix
%% command and will number appendix equations (A1), (A2), etc. The
%% Figure and Table counter will not reset.

\appendix
\section{Degeneracy between mass mixing ratio and gravity} \label{sec:CIA}
To demonstrate the degeneracy between the mass mixing ratio and gravity in the optical depth by molecular line absorption and its irrelevance for the optical depth by collision-induced absorption (CIA), we calculated the spectral models for two cases, the fiducial case and the case where the gravity and mass mixing ratio are doubled. Note that the mass is also doubled to have the same value for the square of the radius, namely the flux. {Additionally, the volume mixing ratios of \mo{H_2} and \mo{He} were adjusted to ensure that the total volume mixing ratio sums to one.}
The parameter values adopted for the fiducial case  {correspond to the median values from the mass-constrained retrieval (see Table~\ref{tab:values})}: $M = 71.47$, $\log_{10} g = 5.01$, $\log_{10} x_\mathrm{H_2O} = -2.85$, $\log_{10} x_\mathrm{CH_4} = -2.97$, $T_0 = 1000.23$, $\alpha = 0.088$, $\mathrm{RV} = 5.96$, and $v \sin{i} = 18.51$
(we ignored \mo{NH_3} absorption).
%, where the unit for each parameter is the same as that used in Table~\ref{tab:params}.

\begin{figure}[ht!]
\plotone{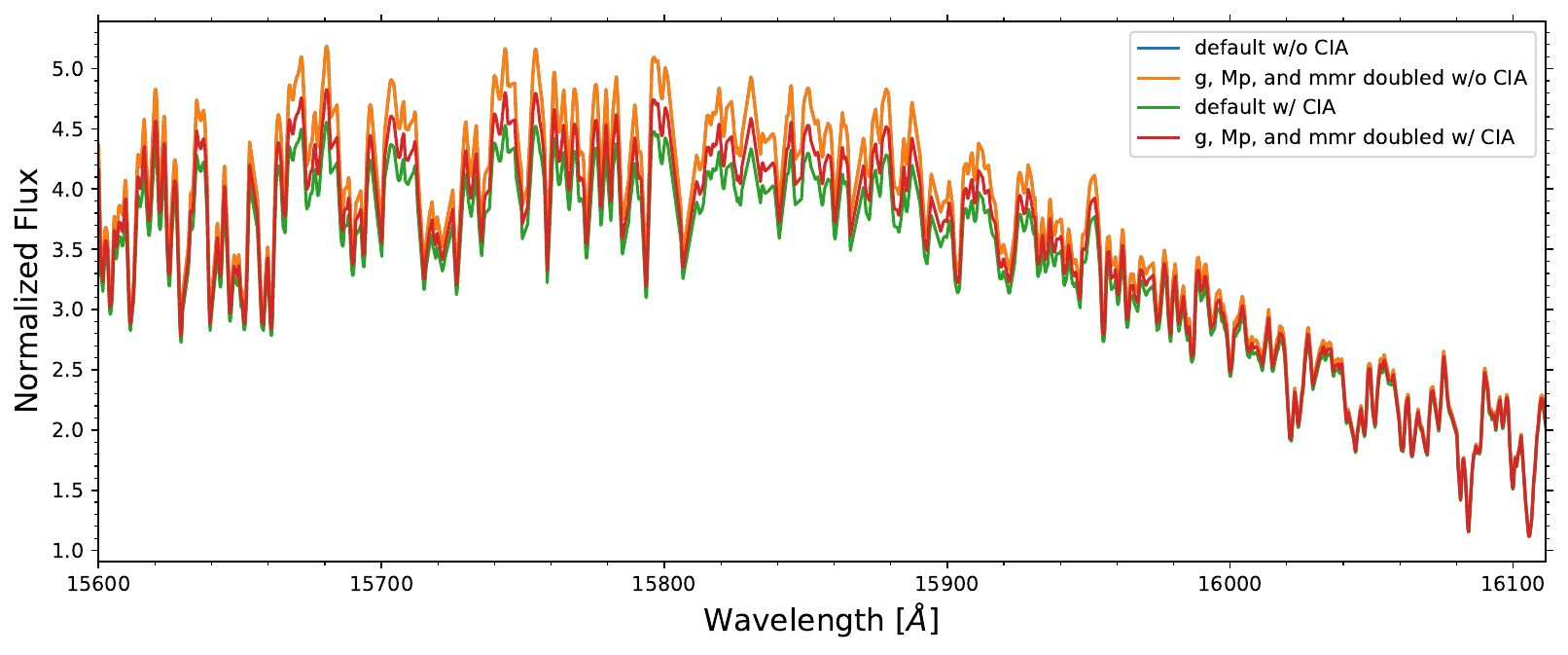}
\caption{Spectral models with and without CIA absorption (green and blue lines, respectively) for the fiducial case compared to those with and without CIA absorption (red and orange lines, respectively) for the case where the gravity and mass mixing ratio are doubled.
Note that the blue line is difficult to see due to the overlap with the orange line.
\label{fig:CIA}}
\end{figure}

In Figure~\ref{fig:CIA}, we show the calculated spectral models. The fiducial models with and without CIA absorption (\mo{H_2}--\mo{H_2}/\mo{H_2}--\mo{He}) are compared to the ``doubled" models with and without CIA absorption. 
It can be seen that when the CIA absorption is ignored, the fiducial and doubled spectral models (blue and orange lines, respectively) coincide with each other.
In contrast, including the CIA absorption yields a difference for the two models (green and red lines, respectively), helping break the degeneracy between the mass mixing ratio and gravity.

%\section{Best-fit model spectra for each IRD spectral order in $H$-band} \label{sec:opt}
%\input{optimization.tex}

%% For this sample we use BibTeX plus aasjournals.bst to generate the
%% the bibliography. The sample631.bib file was populated from ADS. To
%% get the citations to show in the compiled file do the following:
%%
%% pdflatex sample631.tex
%% bibtext sample631
%% pdflatex sample631.tex
%% pdflatex sample631.tex

\section{Impact of spectral resolution on simulated magnitudes}
\label{sec:res}
\begin{figure}[ht!]
\plotone{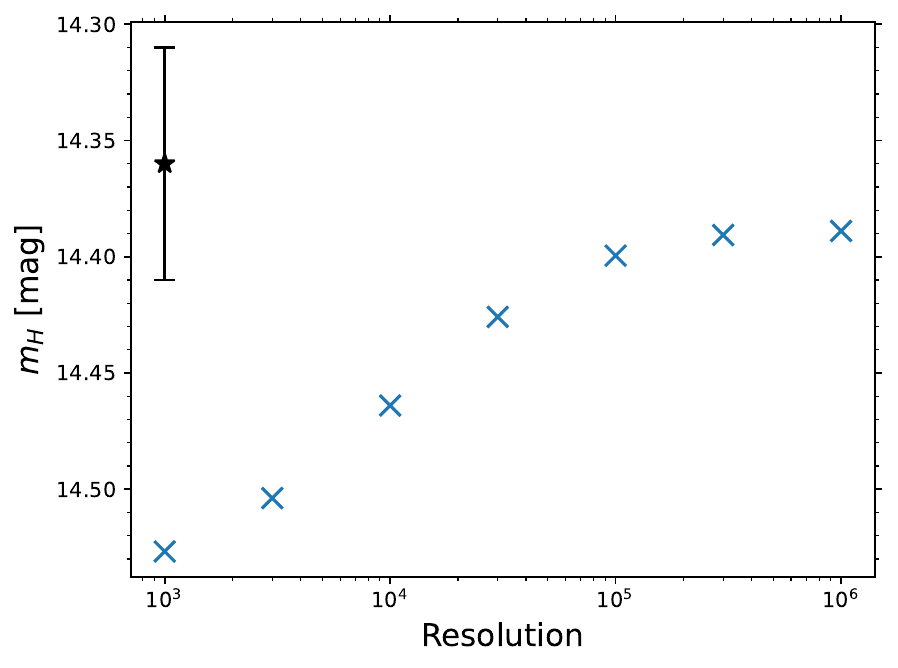}
\caption{Dependence of the simulated $H$-band magnitude on the spectral resolution used in the calculation of the spectrum models. The black star indicates the observed $H$-band magnitude of $14.36 \pm 0.05$~mag \citep{2012ApJ...752...56F} for reference.
\label{fig:Hmag}}
\end{figure}

{Figure~\ref{fig:Hmag} shows how the simulated magnitude depends on the spectral resolution used in the calculation of the spectrum models.
In this demonstration, we adopted the median values from the mass-constrained retrieval (see Table~\ref{tab:values}).
It is clearly evident that the magnitude is systematically underestimated at insufficient spectral resolution.}

\begin{figure}[ht!]
\plotone{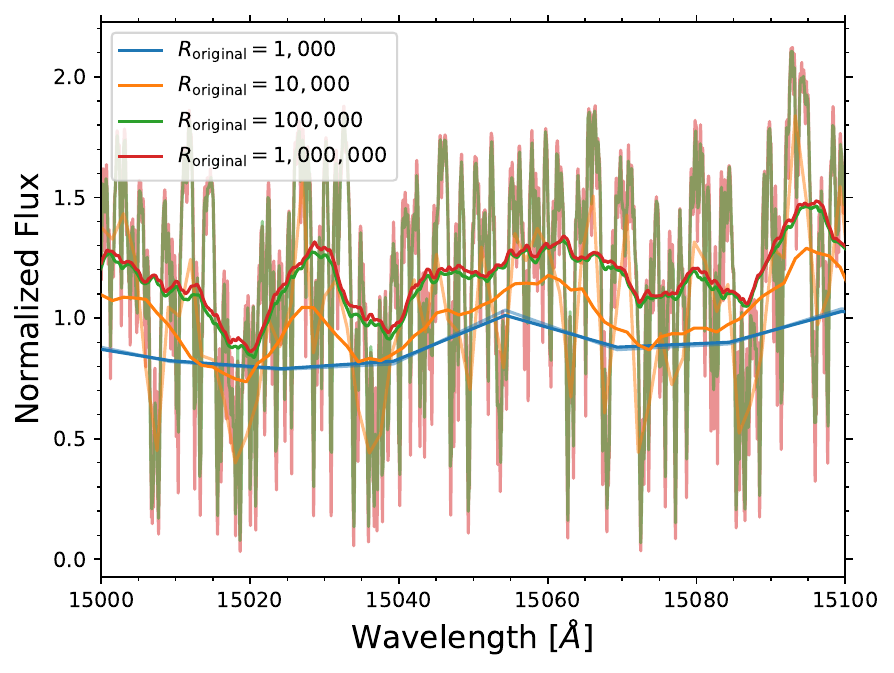}
\caption{Spectrum models degraded to a resolution of $R = 1,000$ (thick lines), based on original models calculated at resolutions of $R = 1,000$ (blue), $10,000$ (orange), $100,000$ (green), and $1,000,000$ (red; thin lines).
\label{fig:Hmag_sp}}
\end{figure}

{This occurs because, although the cross-section retains line intensity regardless of resolution, this conservation does not hold when converting to the emission spectrum, as illustrated in Figure~\ref{fig:Hmag_sp}.
Note that in the spectrum models shown in Fig.~\ref{fig:Hmag_sp}, both $\mathrm{RV}$ and $v \sin{i}$ are set to zero for clarity.
At lower spectral resolutions, the portion of the atmosphere being resolved becomes narrower, causing the line core and wings to probe deeper and higher atmospheric layers, respectively, compared to higher resolutions.}

{Given that flux approximately scales with $T^4$, and assuming a non-inverted temperature--pressure profile where temperature increases with depth, the average flux calculated at lower spectral resolution becomes systematically smaller, as clearly demonstrated by the spectrum models degraded to $R = 1,000$ (thick lines) in Fig.~\ref{fig:Hmag_sp}.
This is because the reduction in flux from the cooler temperatures in the line wings ($T_\mathrm{wing}$) outweighs the increase from the hotter temperatures in the line core ($T_\mathrm{core}$), since $T_\mathrm{wing}^4 > T_\mathrm{core}^4$.}

\bibliography{ads}{}
\bibliographystyle{aasjournal}

%% This command is needed to show the entire author+affiliation list when
%% the collaboration and author truncation commands are used.  It has to
%% go at the end of the manuscript.
%\allauthors

%% Include this line if you are using the \added, \replaced, \deleted
%% commands to see a summary list of all changes at the end of the article.
%\listofchanges

\end{document}